\documentclass[11pt]{article}
\usepackage{amsmath,amssymb,mathrsfs}
\usepackage{graphicx,color}
\usepackage{booktabs}

\newtheorem{definition}{Definition}
\newtheorem{proposition}{Proposition}
\def\Ione{\hbox{\rm 1\kern-3pt l}}
\def\sss{\scriptscriptstyle}
\def\wQ{\mathop{\hbox{$\wedge Q$}}}
\def\9{{\vphantom{1}}}
\def\Gr{\mathop{\text{\rm Gr}}\nolimits}
\def\Span{\mathop\text{\rm Span}\nolimits}
\def\Sym{\mathop\text{\rm Sym}}

\allowdisplaybreaks
   \parskip\medskipamount     \lineskip=0pt
\thicklines

\makeatletter
\@addtoreset{equation}{section}
 \makeatother

\begin{document}
\thispagestyle{empty}
 \noindent
 \begin{center}
{\LARGE\bf Superspace: a Comfortably Vast Algebraic Variety\footnote{Presented at the conference {\em\/Geometric Analysis: Present and Future\/}, Cambridge, MA, August-September 2008.}}\\*[5mm]
{\large\bf T.\,H\"{u}bsch}\\*
{\it Department of Physics \&\ Astronomy,\\[-1mm]
     Howard University, Washington, DC 20059\\[-1mm]
     Department of Physics, Faculty of Natural Sciences\\[-1mm]
     University of Novi Sad, Novi Sad, Serbia\\[-1mm]
    {\small\tt  thubsch@howard.edu}
 }\\[3mm]
{\bf ABSTRACT}\\[3mm]
\parbox{135mm}{\baselineskip=12pt
Supersymmetry has been studied for over three decades by physicists, its superset even longer by mathematicians, and {\em\/superspace\/} has proven to be very useful both conceptually and in facilitating computations. However, the (1)~necessary existence of superspace has been doubted, and its (2)~properties and (3)~applications have not been understood in general. Herein, all doubt is removed from the first of these: superspace must exist. Further study then reveals a perhaps surprising size and algebro-geometric structure of this extension of spacetime.}

\vspace{2mm}\small\sl
{\bfseries Dedicated to Prof.\,Shing-Tung\,Yau}\\[-.5mm]
Many happy returns on your 60$^\text{th}$ birthday!\\[-.5mm]
Boldog 60-ik sz\"ulet\'esnapot!
\end{center}
\begin{flushright}\small\sl
There's a crack in everything;\\[-1mm]
that's how the light comes in.\\[-1mm]
--~Leonard Cohen
\end{flushright}

\section{Introduction}
 \label{s:I}
Super-algebras and many of their special cases have been actively studied by mathematicians much longer than their application in physics, and arguably|in their earliest form|since the work of Saint-Venant and Grassmann~\cite{rGrassmann}, in 1844. Among these, {\em\/supersymmetry algebras\/} have been recognized to be of particular physics interest: they refer to the special cases where the even part of the super-algebra contains the Poincar\'e algebra of symmetries of spacetime $X$, and the inclusion of the odd elements ({\em\/supercharges\/}) $Q_I$, with $I=1,\cdots,N$ extends this to a super-algebra~\cite{rGL71,rR71,rAV,rSSSS1,rKac,r1001,rYM84New,rPW,rWB,rBDW,rYM97Gau,rBK,rMFTS,rDF,rVSV,rARog}.

In view of this history and the extended literature on the subject, it is a little surprising to find an unexpectedly vast {\em\/superspace\/}
\begin{equation}
  \mathscr{S}\!\!_{\sss N}(X) = (\wQ \,(X);
                         Y;Z;\cdots)
 \label{eSNX}
\end{equation}
as a consistent supersymmetric extension of spacetime $X$, with an indefinitely and hierarchically telescoping algebro-geometric structure, only a very small part of which seems to have been used so far.

\subsection{Basic Ideas and Definitions}
A {\em\/supermultiplet\/} $\mathscr{M}$ is a collection of {\em\/component fields\/}: bosonic ($\phi$) and fermionic ($\psi$) functions over a given spacetime $X$, such that the supercharges $Q_I$ map bosons to fermions\footnote{Representations of the Poincar\'e algebra are herein regarded as functions over spacetime, which span representations of the Lorentz subalgebra; its tensorial representations are normally identified as {\em\/bosons\/}, while the spinorial ones are {\em\/fermions\/}.} (and their spacetime derivatives), and fermions to bosons (and their spacetime derivatives) in a system where this transformation is a (super)symmetry.
The $Q_I$-image of any component field is its $I^\text{th}$ (immediate) {\em\/superpartner\/}, and a supermultiplet $\mathscr{M}=(\phi_1,\cdots|\psi_1,\cdots)$ must be {\em\/complete\/}: the superpartner of every component field of $\mathscr{M}$ must also be in $\mathscr{M}$, {\em\/i.e.\/}, must be a linear combination of component fields and their spacetime derivatives. Repeated application of the $Q_I$'s on every component field must satisfy the supersymmetry algebra relations. Of practical interest are supermultiplets with a finite number of component fields|the finite (and, with proper definitions, unitary) representations of supersymmetry.

Supermultiplets in which the component fields are required to satisfy some spacetime differential equation|perhaps so as to ensure the completeness of the $Q$-orbit|are called {\em\/on-shell\/}; if no such requirement is needed or imposed, the supermultiplet is {\em\/off-shell\/}. The latter kind of supermultiplets are indispensable in the quantum theory, by definition, to ensure unobstructed quantum fluctuations of all fields. It is then unsettling that for most supersymmetric theories, mostly those with more than $N=8$ supercharges, no off-shell descriptions are known. The ensuing ``wish list'' of desirable results about supersymmetry could be most easily described by analogy with the well-known example of the $\mathfrak{su}(2)$ algebra:
\par\noindent{\bf Wish-List for Off-Shell Supersymmetry}%
 \hrulefill\vrule height.4pt depth6pt width.4pt
\begin{subequations}\label{e:Task}
\newline\leavevmode\hbox to\parindent{\hss\bf1.~}\ignorespaces
 The complete Hilbert $\mathscr{H}$ space of finite-dimensional unitary representations\footnote{The relation ``$:=$'' is herein used to mean that the left-hand side is defined to equal the right-hand side.}, such as:
\begin{equation}
 \mathfrak{su}(2):~
 \mathscr{H}:=
 \big\{|j,m\rangle,\text{ for each }2j\in\mathbb{Z}\,:\quad
                         |m|\leq j,~ (m{-}m')\in\mathbb{Z};\quad
                         \langle j',m'|j,m\rangle=\delta_{j,j'}\delta_{m,m'}\big\}.
 \label{e1}
\end{equation}
\newline\leavevmode\hbox to\parindent{\hss\bf2.~}\ignorespaces
 The internal tensor product decomposition algorithm within $\mathscr{H}$, such as:
\begin{equation}
  \mathfrak{su}(2):~
  |j',m'\rangle\otimes|j'',m''\rangle = 
  \bigoplus\nolimits_{j=|j'{-}j''|}^{j'{+}j} \big|j,m{:=}(m'{+}m'')\big\rangle.
 \label{e2}
\end{equation}
\newline\leavevmode\hbox to\parindent{\hss\bf3.~}\ignorespaces
 The $\Bbbk$-valued Clebsh-Gordan coefficients in Wish~\#2, such as:
\begin{equation}
  \mathfrak{su}(2):~
  C^{j,m}_{j',m';j'',m''} := \langle j,m|j',m';j'',m''\rangle \in \Bbbk;\quad
  |j',m';j'',m''\rangle:=|j',m'\rangle\otimes|j'',m''\rangle.
 \label{e3}
\end{equation}
\end{subequations}
\vrule height6pt width.4pt\hrulefill\vrule height6pt width.4pt\newline
The analogues of these results are known for all classical Lie algebras, although not in such an explicit and closed form, but as a constructive, iterative algorithm.
For off-shell representations of supersymmetry algebras, such results are sorely lacking.

Indeed, already the first of the above-listed tasks, the complete classification of {\em\/off-shell\/} representations of the $N$-extended supersymmetric extension of the Poincar\'e algebra, akin to~(\ref{e1}) for $\mathfrak{su}(2)$, remains an open problem: even in the simple case of {\em\/worldline\/} supersymmetry, where spacetime is reduced to $\mathbb{R}^1$ of time, a complete classification is {\em\/only now emerging\/}~\cite{r6-1,r6--1,r6-3c,r6-3,r6-3.2}! These efforts successfully employ a synergy with graph theory and error-correcting codes, and have already uncovered a surprisingly large combinatorial complexity: over a trillion inequivalent supermultiplets are expected to exist for supersymmetry with $N\leq32$ generators|and this is {\em\/before\/} ``tensoring and linear algebra''\footnote{It is standard that all finite-dimensional unitary representations of classical Lie algebras may be constructed from a single, {\em\/fundamental\/} representation, by ``tensoring and linear algebra'': One iterates the procedure of taking tensor products of this fundamental representation with itself, and then identifying kernels or cokernels of proper homomorphisms between such tensor products, typically including symmetrization or contraction with any invariants of the given Lie algebra. For example, the only $\mathfrak{su}(n)$-invariant is the Levi-Civita volume-form $\varepsilon$, while its $\mathfrak{so}(n)$ subalgebra also has the $n{\times}n$ Kronecker $\delta$ as the metric; $\mathfrak{sp}(2n)$ instead has the Levi-Civita volume-form $\varepsilon$ and the symplectic 2-form $\Omega$ as invariants, {\em\/etc\/}.}.
\newline\centerline{|~$\star$~|}

Rooted in the synergy between algebra and geometry is the approach wherein spacetime is extended to {\em\/superspace\/}, dually to the Poincar\'e algebra being extended by supersymmetry. Supermultiplets are then understood as generalized functions over superspace, called {\em\/superfields\/}. Action functionals governing the dynamics\footnote{Hamilton's action functional, $\mathcal{S}[\phi,\psi]$, is essential to all physics: classical physics restricts $\phi,\psi$ so as to minimize $\mathcal{S}[\phi,\psi]$; in quantum theory, moments of the path integral $\int\! D[\phi,\psi]\,\exp\{i\mathcal{S}[\phi,\psi]/\hbar\}$ determine the probabilities of all processes.} of the superfields in any particular model of physics interest are then expressible as superspace integrals of functional expressions constructed from superfields and their super-derivatives, thus extending in a natural way the well-understood Lagrangian/Hamiltonian approach. 

This {\em\/superspace approach\/} has been successfully employed for many of the simpler models~\cite{rSSSS1,r1001,rWB,rBK} and has been well studied in both the physics and mathematics literature. Nevertheless, it seems to fall short in several important aspects of physics interest, and foremost in cases where the total number of supersymmetry generators exceeds $N=8$. In particular, the unexpectedly large number of supermultiplets~\cite{r6-3,r6-3.2} indicates a possibly serious mismatch with the rather more modest expectations and results stemming from superspace practice so far.

The main goal herein is then to revisit the meaning and structure of superspace, and it turns out that this supersymmetric extension of spacetime~(\ref{eSNX}) is much larger than what has been so far known and used in the literature, both in mathematics~\cite{rYM84New,rYM97Gau} and physics~\cite{r1001,rWB,rPW,rBK}. The remainder of this section presents a few well-established facts about this traditional superspace. Section~\ref{s:2Many} outlines the emerging classification of off-shell representations of $N$-extended worldline supersymmetry, and Section~\ref{s:SSpace} then presents an explicit, iterative construction of an unexpectedly large superspace for worldline supersymmetry with $N$ generators, with Section~\ref{s:Higher} collecting a few remarks about generalizations to higher-dimensional spacetime and possible applications of this novel superspace.

\subsection{The Traditional Superspace}
 \label{s:XTheta}
Recall a few rigorous definitions of superspace and supermanifolds~\cite{rYM84New,rYM97Gau}:
\begin{definition}[Manin]\label{d:SS}
A {\em\/superspace\/} is defined to be a pair $(M,\mathcal{O}_M)$ consisting of a topological space $M$ and a sheaf of super-commutative rings $\mathcal{O}_M$ on it such that the stalk $\mathcal{O}_x=\mathcal{O}_{M,x}$ at any point $x\in M$ is a local ring.
\end{definition}
Denote by $\mathcal{O}_M^e$ the even and by $\mathcal{O}_M^o$ the odd part of $\mathcal{O}_M$, define $J_M:=\mathcal{O}_M^o+(\mathcal{O}_M^o)^2$, a sheaf of ideals in $\mathcal{O}_M$, and set
\begin{equation}
 \Gr_i\mathcal{O}_M:=J_M^i/J_M^{i+1}.
 \label{eGr}
\end{equation}
Then $\Gr_0\mathcal{O}_M=\mathcal{O}_M/J_M$ is a sheaf of rings, $M_\text{rd}:=(M,\Gr_0\mathcal{O}_M)$ is purely even superspace, and|for {\em\/supermanifolds\/}|equals $M_\text{red}:=(M,\mathcal{O}_M/\mathcal{N})$, where $\mathcal{N}$ is the sheaf of all nilpotents in the structure sheaf. A useful object is the ringed space
\begin{equation}
  \Gr M:=(M,\Gr\mathcal{O}_M)=(M,\bigoplus\nolimits_{i\geq0}\Gr_i\,M),
\end{equation}
which may also be regarded as a superspace, with its $\mathbb{Z}_2$ grading being the modulo-2 reduction of the natural $\mathbb{Z}$-grading.
\begin{definition}[Manin]\label{d:SM}
A {\em\/supermanifold\/} is a superspace $(M,\mathcal{O}_M)$ such that $M_\text{rd}$ is purely even manifold and $\mathcal{O}_M$ is locally isomorphic to $\Gr\mathcal{O}_M$, which, in turn, is isomorphic to the Grassmann algebra of the locally free sheaf over $\mathcal{O}_M/J_M$ of finite rank $J_M/J_M^2$.
\end{definition}

Coupled with the Poincar\'e-Birkhoff-Witt type theorems for super-algebras and the fact the supersymmetry algebra|and in particular its odd part (generated by $N$ supercharges $Q_I$)|is supposed to act nontrivially on the superspace and functions thereof, one expects $\mathcal{O}_M$ to be modeled on the exterior algebra $\wQ :=\bigwedge\Span(Q_1,\cdots,Q_N)$. Indeed, since the original introduction of superspace into the physics literature~\cite{rSSSS1}, it has been modeled by augmenting the usual (bosonic, commuting) coordinates of spacetime with the unusual (fermionic, anticommuting) coordinates, $\theta$, one for each generator of supersymmetry. Superfields are then easily expanded over such $\theta$-monomials,
\begin{equation}
  \Phi(x,\theta) = \phi_{[0]}(x) + \theta{\cdot}\psi_{[1]}(x)
 + \wedge^2\theta{\cdot}\phi_{[2]}(x) + \wedge^3\theta{\cdot}\psi_{[3]}(x) + \dots
 \label{eSFX}
\end{equation}
defining the corresponding supermultiplet
$\mathscr{M}_\Phi=(\phi_{[0]},\phi_{[2]},\cdots|\psi_{[1]},\psi_{[3]},\cdots)$, which obviously terminates with the ``top'' component field, occurring with $\wedge^N\theta$ in the expansion. Just as the linear momenta generate translations in spacetime $X$, are therefore representable as derivatives with respect to spacetime coordinates $^ix$, and so are identifiable with tangent vectors in $T_X$, the supersymmetry generators $Q_I$ may be represented by ($X$-twisted) derivatives with respect to $\theta^I$, and so are identifiable as tangent to the fermionic (odd) extension of spacetime. This canonical duality is reflected in the {\em\/canonical super-commutation relations\/}\footnote{For the classical Hamiltonian formalism, the commutator is replaced with the $i\hbar$-multiple of the Poisson bracket.}:
\begin{equation}
  \big[P_j,x^k\,\big]=i\hbar\,\delta_j{}^k,\qquad
  \big\{Q_I,\theta^J\,\big\} = \delta_I{}^J.
 \label{eCCRX}
\end{equation}

So, while the physics literature refers to $\Span(x^0,x^1\cdots|\theta^1,\theta^2\cdots)$ as superspace, the definitions~\ref{d:SS}--\ref{d:SM} and the PBW-type theorems pair with this a sheaf modeled on $\mathcal{O}_X[\wQ]$, {\em\/i.e.\/}, formal polynomials in elements of $\wedge\theta$ (coordinates on $\wQ$), and with coefficients in $\mathcal{O}_X$, the sheaf of functions of a desired type (smooth, analytic, holomorphic\dots) over the spacetime $X$. Both however agree that superfields|the basic objects from which the finite-dimensional unitary representations of supersymmetry ought to be constructed by ``tensoring and linear algebra''|are then formal expansions of the form~(\ref{eSFX}).
\newline\centerline{|~$\star$~|}

For the rest of this paper, until Section~\ref{s:Higher}, reduce spacetime to the {\em\/worldline\/}, $\mathbb{R}^1_\tau\subset X$, and replace spacetime coordinates $x\to\tau$, which may be identified with the {\em\/proper time\/}. So, $\tau$ is an observer-preferred real-valued function over the spacetime, $\tau:X\to\mathbb{R}^1$. This reduces the supersymmetry algebra (without central charges) to:
\begin{equation}
 \begin{gathered}
  \{Q_I,Q_J\} = 2\,\delta_{IJ}\,H,\qquad
  [H,Q_I]=0,\qquad I,J=1,\cdots,N,\\
  (Q_I)^\dagger=Q_I,\qquad (H)^\dagger = H,
 \end{gathered}
 \label{eSuSy}
\end{equation}
where $H$ is the worldline Hamiltonian, identifiable with $i\hbar\partial_\tau$, and $Q_I$ is the $I^\text{th}$ supercharge. Notably, this avoids all technical and notational difficulties related to the Lorentz symmetry in actual spacetimes. However, those considerations can be temporarily treated as ``internal'' symmetries, unrelated to spacetime, and can be included subsequently in the reverse of the dimensional reduction process, known as dimensional oxidization~\cite{rGR0}.

The $N$-extended worldline supersymmetry algebra~(\ref{eSuSy}) garners physical interest through three separate and logically independent applications:
\begin{enumerate}\itemsep=-3pt\vspace{-3mm}
 \item the dimensional reduction of any supersymmetric theory in ``actual'' spacetime, such as supersymmetric Yang-Mills gauge theories, the supersymmetric Standard Model of particle physics, {\em\/etc.\/};
 \item the {\em\/underlying\/} description or dimensional reduction thereof, in multi-layered physical theories such as the worldsheet description of superstring theory, or the matrix version of $M$-theory;
 \item the induced supersymmetry in the Hilbert space of a supersymmetric theory, in the  Schr\"odinger picture, where $H$ and $Q_I$ are expressed in terms of particle state creation and annihilation operators.
\end{enumerate}\vspace{-3mm}
Although not limited in principle, $N\leq32$ seems to suffice in all currently known fundamental physics.

A few remarks are in order about the algebra~(\ref{eSuSy}) and the only other ``physics input'':
\begin{equation}
  [H,\tau]=i\hbar,
  \quad\text{so}\quad
  H=i\hbar\partial_\tau~\text{on functions of $\tau$}.
 \label{eCCR}
\end{equation}
\begin{itemize}\itemsep=-3pt
 \item The supersymmetry algebra has a $\tfrac12\mathbb{Z}$-grading, called {\em\/engineering dimension\/}, defined by specifying
  \begin{equation}
    [\tau]=-1,\qquad
    \buildrel{\smash{(\ref{eCCR})}}\over\Longrightarrow\qquad
    [H]=1,\quad\Rightarrow\quad[Q]=\tfrac12.
   \label{eED}
  \end{equation}
In the unit system where $\hbar$ and $c$ are two of the three basic (and unwritten) units, the engineering dimension is the exponent of the mass/energy unit.

 \item Supermultiplets may be constructed, in a Fock-space manner, as complete chains of superpartners\footnote{Herein, ``$\wedge^kQ\,(f)$'' denotes the result of applying $\wedge^kQ$ on $f$, not the $k$-fold exterior power of $Q(f)$.}:
  \begin{equation}
   \big\{\phi_{[0]},~ \psi_{[1]}:=Q(\phi_{[0]}),~
            \phi_{[2]}:=Q(\psi_{[1]})=\wedge^2Q\,(\phi_{[0]}), \cdots ,
             \wedge^NQ\,(\phi_{[0]})\,\big\}
   =: \wQ\,(\phi_{[0]}),
   \label{ePBW}
  \end{equation}
reinforcing the PBW-based expectations that the structure sheaf of a superspace, in definition~\ref{d:SS}, is modeled on $\mathcal{O}_X[\wQ ]$. Note: if $Q(\phi)=\psi$ for some $\phi$ and $\psi$, then $Q(\psi)=H(\phi)=i\hbar\dot\phi$, with $\dot\phi:=(\partial_\tau\phi)$. Note that the supermultiplet~(\ref{ePBW}) corresponds precisely to the superfield~(\ref{eSFX}).

 \item Combining the above two points and Eq.~(\ref{eSFX}) yields:
  \begin{equation}
   [\phi_{[2k]}]=[\phi_{[0]}]+k,\quad
   [\psi_{[2k+1]}]=[\phi_{[0]}]+k+\tfrac12,\qquad
   [\theta]=-\tfrac12.
   \label{eEDq}
  \end{equation}

 \item Up to the over-all, additive $[\phi_{[0]}]$, the ``mod~2'' reduction of the {\em\/double\/} of the engineering dimension $\tfrac12\mathbb{Z}$-grading corresponds precisely to the $\mathbb{Z}_2$-grading required in super-algebras.
\end{itemize}

\section{Off-Shell Worldline Supermultiplets}
 \label{s:2Many}
A sequence of studies~\cite{rGR-1,rGR0,rGR1,rGR2,rGR3,rGLPR,rGLP} and then~\cite{rA,r6-1,r6--1,r6-3c,r6-3,r6-3.2} forged a novel approach to the problem of classifying off-shell supermultiplets of the ``$N$-extended worldline supersymmetry algebra without central charges''~(\ref{eSuSy}). Notably, this approach eschews direct recourse to superspace although results are verified for consistency where possible. Instead, it employs graph theory, and turns out to also involve error-correcting codes.
 Application of these techniques to concrete and previously unsolved problems in supersymmetric physics was demonstrated in Ref.~\cite{r6-2,r6-3a,r6-4,r6-7a,r6-4.2}.

\subsection{Adinkraic Supermultiplets}
 \label{s:A}
Focus on supermultiplets in which the $Q$-image of every single component field is again a single component field, so that the algebra~(\ref{eSuSy}) implies~\cite{r6-3}:
\begin{definition}\label{dAd}\vspace{-2mm}
A supermultiplet $\mathscr{M}$ is {\bfseries\/adinkraic\/} if it admits a basis, $(\phi_1,\cdots,\phi_m\,|\,\psi_1,\cdots,\psi_m)$,  of component fields such that each $Q_I\in\{Q_1,\cdots,Q_N\}$ acts upon each $\phi_A\in\{\phi_1,\cdots,\phi_m\}$ so as to produce:\vspace{-4mm}
\begin{subequations}\label{eQbf}
 \begin{align}
 Q_I \, \phi_A(\tau) &= c\,\partial_\tau^{\lambda}\, \psi_B(\tau),
 \quad\text{where}\quad
  c=\pm1,\quad \lambda=0,1,\quad\psi_B\in\{\psi_1,\cdots,\psi_m\},
   \label{eQb}\\[-2mm]
\intertext{and the right-hand side choices depend on $I$ and $A$. In turn, this $Q_I$ acting on this $\psi_B$ produces:\vspace{-2mm}}
 Q_I \, \psi_B(\tau) &= \frac{i}{c}\,\partial_\tau^{1-\lambda}\, \phi_A(\tau),\label{eQf}
\end{align}
\end{subequations}
and the pair of formulae~{\rm(\ref{eQbf})} exhausts the action of each $Q_I$ upon each component field.
\end{definition}

The structure of an adinkraic supermultiplet may be faithfully depicted by an {\em\/Adinkra\/}:
 ({\small\bf1})~Assign a node to every component field: white for bosons and black for fermions.
 ({\small\bf2})~Draw an edge in the $I^\text{th}$ color from node $v_1$ to node $v_2$ precisely if the component field $F_2$ of $v_2$ is the $Q_I$-image of the component field $F_1$ of $v_1$ and $[F_2]=[F_1]+\frac12$.
 ({\small\bf3})~An edge is drawn solid if $c=+1$ in Eqs.~(\ref{eQb})--(\ref{eQf}), and dashed if $c=-1$.
 See Table~\ref{t:A} for a dictionary.
\begin{table}[ht]
  \centering\setlength{\unitlength}{1mm}
  \begin{tabular}{@{} cc|cc @{}}
    \makebox[15mm]{\bf Adinkra} & \makebox[40mm]{\bf\boldmath$Q$-action} 
  & \makebox[15mm]{\bf Adinkra} & \makebox[40mm]{\bf\boldmath$Q$-action} \\ 
    \hline
    \begin{picture}(5,9)(0,5)
     \put(0,0){\includegraphics[height=11mm]{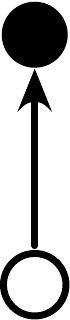}}
     \put(3,0){\scriptsize$A$}
     \put(3,9){\scriptsize$B$}
     \put(-1,4){\scriptsize$I$}
    \end{picture}\vrule depth4mm width0mm
     & $Q_I\begin{bmatrix}\psi_B\\\phi_A\end{bmatrix}
           =\begin{bmatrix}i\dot\phi_A\\\psi_B\end{bmatrix}$
  & \begin{picture}(5,9)(0,5)
     \put(0,0){\includegraphics[height=11mm]{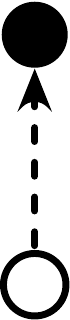}}
     \put(3,0){\scriptsize$A$}
     \put(3,9){\scriptsize$B$}
     \put(-1,4){\scriptsize$I$}
    \end{picture}\vrule depth4mm width0mm
     & $Q_I\begin{bmatrix}\psi_B\\\phi_A\end{bmatrix}
           =\begin{bmatrix}-i\dot\phi_A\\-\psi_B\end{bmatrix}$ \\[5mm]
    \hline
    \begin{picture}(5,9)(0,5)
     \put(0,0){\includegraphics[height=11mm]{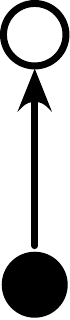}}
     \put(3,0){\scriptsize$B$}
     \put(3,9){\scriptsize$A$}
     \put(-1,4){\scriptsize$I$}
    \end{picture}\vrule depth4mm width0mm
     &  $Q_I\begin{bmatrix}\phi_A\\\psi_B\end{bmatrix}
           =\begin{bmatrix}\dot\psi_B\\i\phi_A\end{bmatrix}$
  & \begin{picture}(5,9)(0,5)
     \put(0,0){\includegraphics[height=11mm]{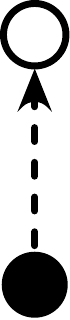}}
     \put(3,0){\scriptsize$B$}
     \put(3,9){\scriptsize$A$}
     \put(-1,4){\scriptsize$I$}
    \end{picture}\vrule depth4mm width0mm
     &  $Q_I\begin{bmatrix}\phi_A\\\psi_B\end{bmatrix}
           =\begin{bmatrix}-\dot\psi_B\\-i\phi_A\end{bmatrix}$ \\[5mm]
    \hline
  \multicolumn{4}{l}{\vrule height3.5ex width0pt\parbox{120mm}{\small The edges are here labeled by the variable index $I$; for any fixed $I$, each corresponding edge is drawn in the $I^{\text{th}}$ color.}}
  \end{tabular}
  \caption{\baselineskip=12pt
  The correspondences between the Adinkra components and supersymmetry transformation formulae~(\ref{eQb})--(\ref{eQf}):
    vertices\,$\leftrightarrow$\,component fields;
    vertex color\,$\leftrightarrow$\,fermion/boson;
    edge color/index\,$\leftrightarrow$\,$Q_I$;
    edge dashed\,$\leftrightarrow$\,$c=-1$; and
    orientation\,$\leftrightarrow$\,placement of $\partial_\tau$.
    They apply to all $\phi_A,\psi_B$ within a supermultiplet and all $Q_I$-transformations amongst them.}
  \label{t:A}
\end{table}
For clarity, we dispense with the arrows on the edges, but position the nodes so that all edges are oriented upward, and the height at which a node is placed is proportional to the engineering dimension of the corresponding component field~\cite{r6-1}.

The ``dictionary'' in Table~\ref{t:A} provides a precise rules for the $Q$-action within the supermultiplet:
\begin{equation}\setlength{\unitlength}{1mm}
 \vcenter{\hbox{\hss\begin{picture}(8,12)
                     \put(0,0){\includegraphics[height=13mm]{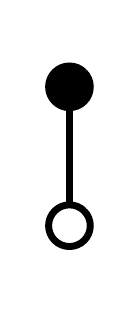}}
                     \put(-2,2){\small$\phi$}
                     \put(-2,10){\small$\psi$}
                    \end{picture}\hss}}
  \Leftrightarrow~
 \left\{~\begin{aligned}
          Q\,\psi &= i\dot\phi,\\
          Q\,\phi &= \psi;
         \end{aligned}\right.
 \qquad\qquad
 \vcenter{\hbox{\hss\begin{picture}(16,12)
                     \put(0,-4){\includegraphics[height=21mm]{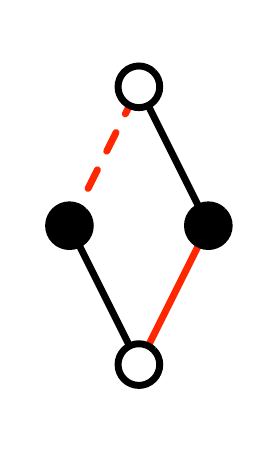}}
                     \put(2,-2){\small$\phi$}
                     \put(-2,7){\small$\psi_1$}
                     \put(11,7){\small$\psi_2$}
                     \put(2,13){\small$F$}
                    \end{picture}\hss}}
 \Leftrightarrow~
 \left\{~\begin{aligned}
          Q_I\,F &= \varepsilon_{IJ}\delta^{JK}\,\dot\psi_K,\\
          Q_I\,\psi_J &= i\delta_{IJ}\dot\phi + i\varepsilon_{IJ}F,\\
          Q_I\,\phi &= \psi_I;
         \end{aligned}\right.
 \qquad\text{\it etc.}
\end{equation}
Two supermultiplets are regarded as inequivalent if it is not possible to transform one into the other by:
\begin{enumerate}\itemsep=-3pt\vspace{-3mm}
 \item a linear combination of the component fields and their $\tau$-derivatives, {\em\/i.e.\/}, by a ``{\em\/field redefinition\/},''
 \item a linear combination of the $Q_1,\cdots,Q_N$.
\end{enumerate}\vspace{-3mm}
The latter transformation is ``outer'' in the sense that in a system involving two separate supermultiplets, $\mathscr{M}_1$ and $\mathscr{M}_2$, a permutation of the $Q_1,\cdots,Q_N$ necessarily affects both supermultiplets. This allows for the interesting possibility: Let the $Q$-action on $\mathscr{M}_2$ be obtainable from the one on $\mathscr{M}_1$ by a linear combination, $\ell_Q$, of the $Q_1,\cdots,Q_N$: $\ell_Q(\mathscr{M}_1)=\mathscr{M}_2$. Then, any action functional involving only one of the two supermultiplets can necessarily be obtained using the other: $\mathcal{S}[\mathscr{M}_1]=\mathcal{S}[\ell_Q(\mathscr{M}_2)]=\ell_Q(\mathcal{S}[\mathscr{M}_2])=\mathcal{S}'[\mathscr{M}_2]$.
 However,
\begin{definition}\label{d:UD}
Two supermultiplets, $\mathscr{M}_1$ and $\mathscr{M}_2$, are {\bfseries\/usefully distinct\/} if a supersymmetric action functional $\mathcal{S}[\mathscr{M}_1,\mathscr{M}_2]$ exists, which could not have been constructed with two copies of either $\mathscr{M}_1$ or $\mathscr{M}_2$. An action functional $\mathcal{S}$ is supersymmetric if $Q(\mathcal{S})=\int{\rm d}\tau\,(\partial_\tau\mathcal{K})$, for some $\mathcal{K}$.
\end{definition}
This {\em\/useful distinctness\/} is a generalization of the quality exemplified by the chiral and twisted-chiral superfields in $(1,1)$-dimensional spacetime and $(2,2)$-supersymmetry~\cite{rGHR}. The definition~\ref{d:UD} however is at once both more general and more subtle: The chiral and twisted-chiral supermultiplets have a distinct {\em\/dashed chromotopology\/}~\cite{r6-3}. For two supermultiplets to be usefully distinct, such a topological distinction is {\em\/a priori\/} not necessary: Even if $\mathscr{M}_2=\ell_Q(\mathscr{M}_1)$, $\mathscr{M}_1$ and $\mathscr{M}_2$ may nevertheless be usefully distinct since it may not be possible to redefine $\{Q_1,\cdots,Q_N\}$ so as to turn $\mathscr{M}_2$ into $\mathscr{M}_1$ without simultaneously transforming $\mathscr{M}_1$ into something else within a given action functional $\mathcal{S}[\mathscr{M}_1,\mathscr{M}_1]$.

It thus behooves to distinguish between equivalences that require $Q$-redefinition, and those that do not.

\subsection{Various Hangings}
All supermultiplets~(\ref{ePBW}) represented by superfields~(\ref{eSFX}) are adinkraic: the Adinkras of the five with the lowest number of supercharges are:
\begin{equation}\setlength{\unitlength}{1mm}
 \vcenter{\hbox{\hss
  \begin{picture}(150,35)
   \put(0,.5){\includegraphics[height=13.5mm]{Pix/xN1B.pdf}}
    \put(-2,15){\small$N=1$}
   \put(12,.5){\includegraphics[height=19.25mm]{Pix/xN2B.pdf}}
    \put(12.5,20){\small$N=2$}
   \put(29,.5){\includegraphics[height=25mm]{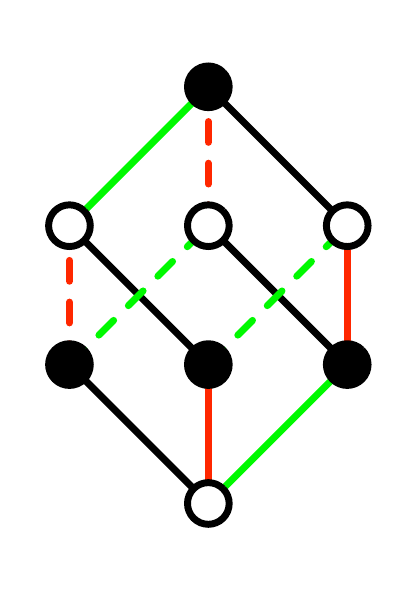}}
    \put(32,25){\small$N=3$}
   \put(52,0){\includegraphics[height=32mm]{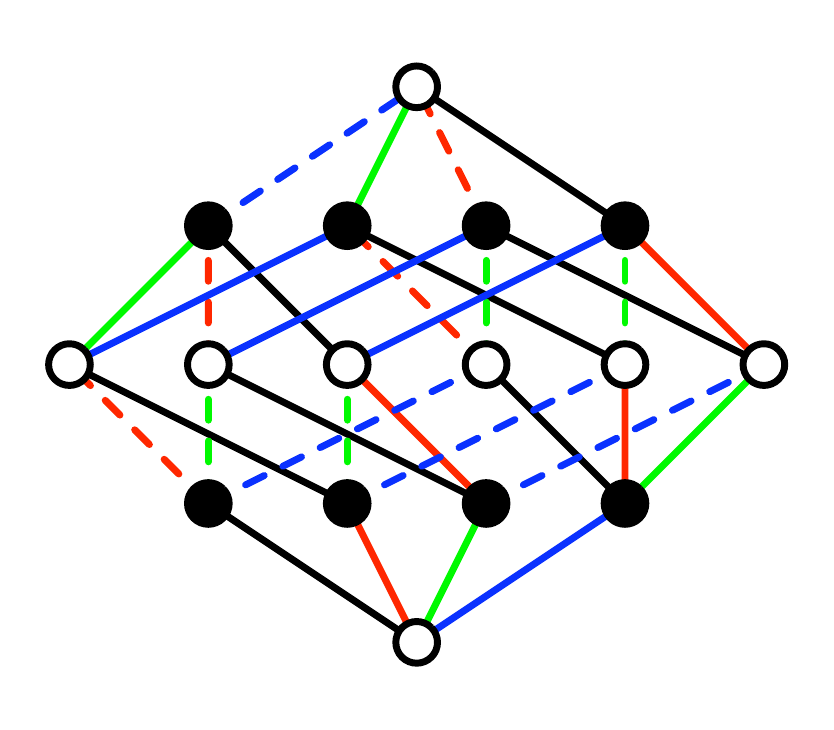}}
    \put(65,31){\small$N=4$}
   \put(92,0){\includegraphics[height=38mm]{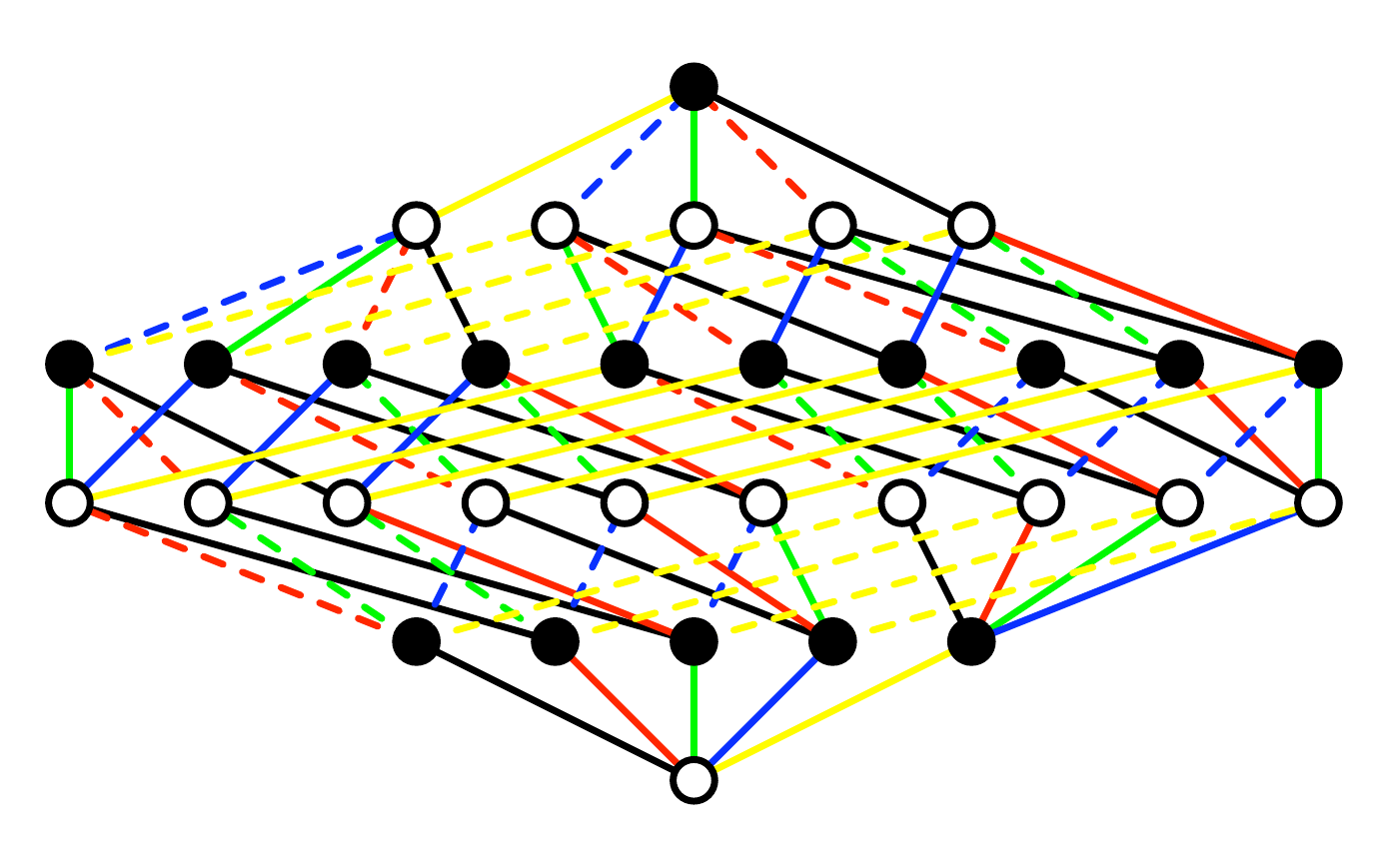}}
    \put(100,32){\small$N=5$}
  \end{picture}\hss}}
 \qquad\text{\it etc.}
 \label{eA1-5}
\end{equation}
The judicious choice of dashed edges ensures that every quadrangle contains edges of two alternating colors and an odd number of dashed edges, reflecting the anticommutativity $\{Q_I,Q_J\}=0$ for $I\neq J$.

This is (by far) not all: many adinkraic supermultiplets do not conform to the expansion~(\ref{eSFX}):
\begin{equation}\setlength{\unitlength}{1mm}
 \vcenter{\hbox{\hss
  \begin{picture}(160,22)
   \put(10,.5){\includegraphics[height=13.5mm]{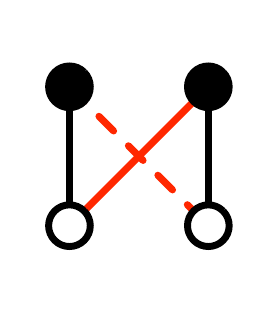}}
    \put(11,14){\small$N=2$}
   \put(27,.5){\includegraphics[height=19.25mm]{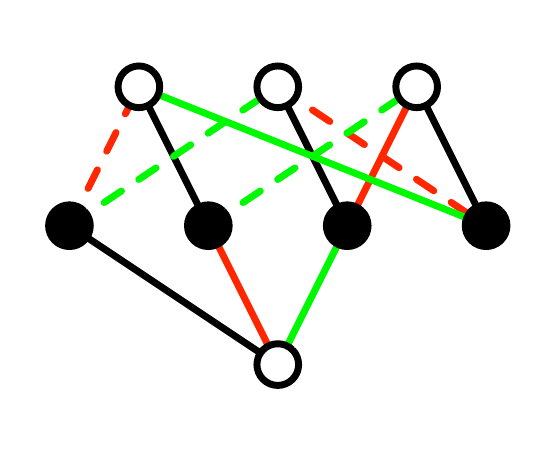}}
   \put(50,.5){\includegraphics[height=19.25mm]{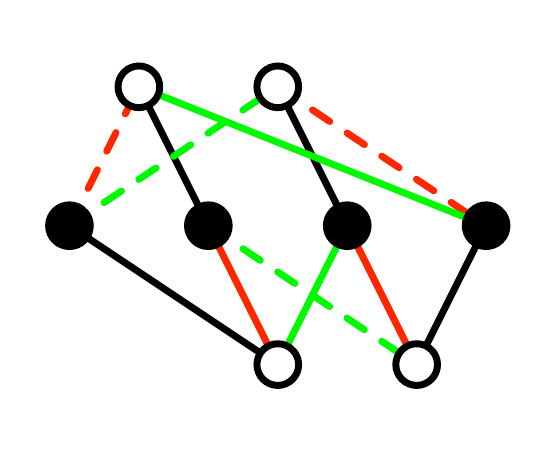}}
    \put(46,19){\small$N=3$}
    \put(73,10){\dots}
   \put(85,0){\includegraphics[height=20mm]{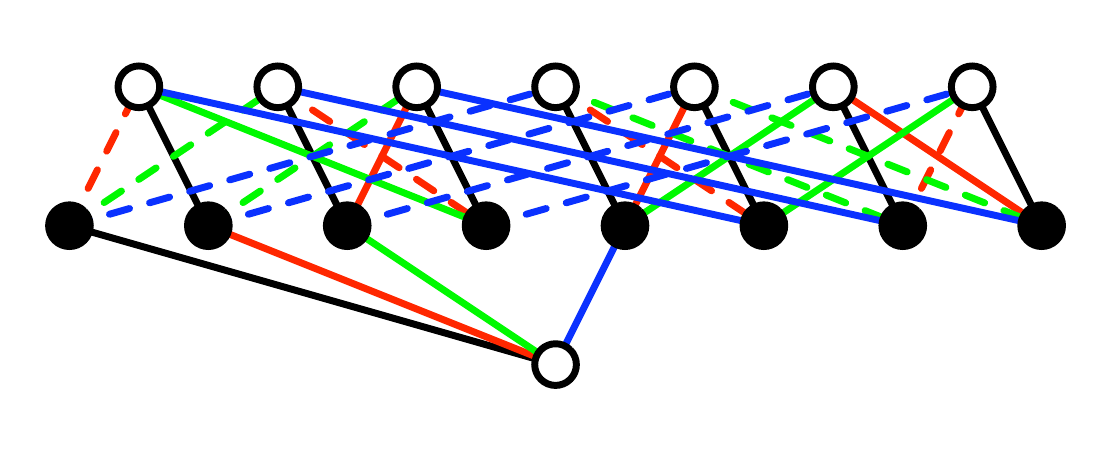}}
    \put(105,19){\small$N=4$}
    \put(134,10){\dots}
    \put(145,19){\small$N>4$}
    \put(147,10){\dots}
  \end{picture}\hss}}
 \label{eNSF}
\end{equation}
These Adinkras differ from those in the sequence~(\ref{eA1-5}) in that some (but not all!) nodes have a different height assignment. Correspondingly, in these supermultiplets some (but not all) of the component fields have an engineering dimension that does not conform to the expansion~(\ref{eSFX}). In comparison, the Adinkras and supermultiplets in the sequence~(\ref{eA1-5}) and the supermultiplets~(\ref{eSFX}) are called {\em\/extended\/}.

It is evident that the number of Adinkras and adinkraic supermultiplets which do not conform to the expansion~(\ref{eSFX})---and so are not maximally extended as those in the sequence~(\ref{eA1-5}) are---grows combinatorially with $N$. As a simple illustration, consider:
\begin{equation}\setlength{\unitlength}{1mm}
 \vcenter{\hbox{\hss\begin{picture}(160,45)(0,-25)
                     \put(0,7){$N=2$:}
                     \put(14,0){\includegraphics[height=22mm]{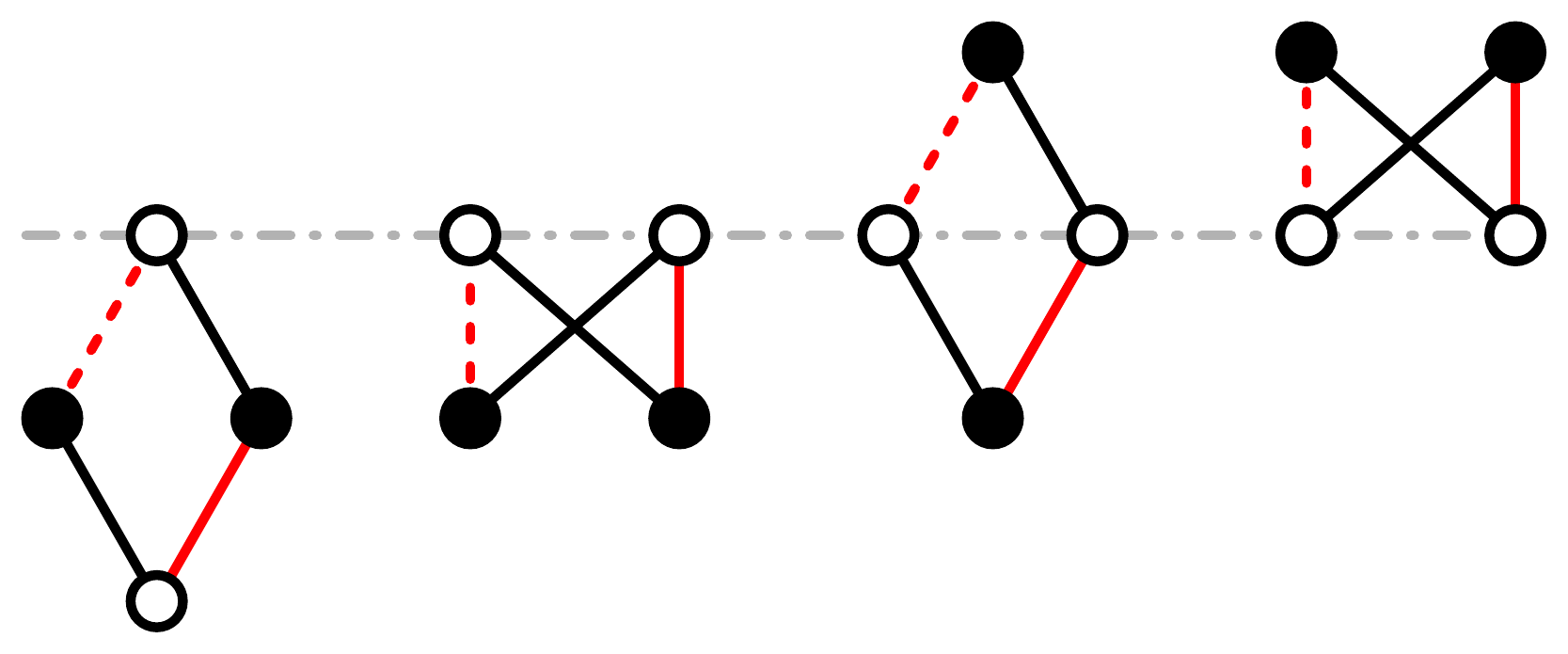}}
                     \put(0,-12){$N=3$:}
                     \put(14,-25){\includegraphics[height=32mm]{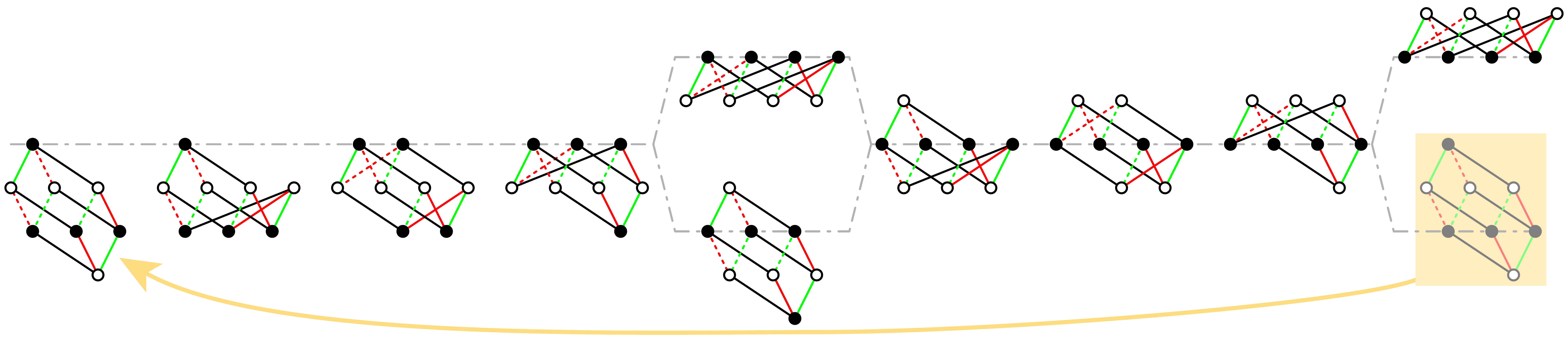}}
                    \end{picture}\hss}}
 \label{eN2+3}
\end{equation}
These sequences have been obtained by systematically raising a node at a time, whereby the corresponding component field is replaced by its $\tau$-derivative. They are cyclic in the sense that sooner or later one obtains a total $\tau$-derivative of a supermultiplet already in the sequence|such as the one highlighted in the $N=3$ sequence; the Adinkra of the latter is identical to the starting one, but with every node two levels higher\footnote{All nodes of all the Adinkras in the sequences~(\ref{eN2+3}) were drawn at constant heights, except for the two instances in the $N=3$ sequence, where there are two distinct ways to raise a node, when the resulting Adinkras are stacked above each other. To guide the eye, a reference  light gray dot-and-dash line is drawn to indicate a reference height (engineering dimension).}.

Order, however, does turn up in this combinatorial complexity of different ``hangings'' of any given Adinkra:
 Theorems~5.1, 7.6 and their Corollaries in Ref.~\cite{r6-1} prove that each adinkraic supermultiplet the Adinkra of which has the same {\em\/topology\/}\footnote{Roughly, the topology of an Adinkra is its 1-skeleton, and specifies the connectivity of the nodes. Including also the node- and vertex-coloring information defines {\em\/chromotopology\/}; for a precise statement, see Refs.~\cite{r6-1,r6-3}.} as one from the sequence~(\ref{eA1-5}) is its variation in ``hanging,'' and may be represented by a {\em\/constrained superfield multiplet\/}, {\em\/i.e.\/}, in terms of superfields. For example:
\begin{align}\setlength{\unitlength}{1mm}
 \text{given a pair,}\quad(\mathbb{A,B})~&\Leftrightarrow
  \vcenter{\hbox{\hss\includegraphics[height=15mm]{Pix/xN2B.pdf}\hss}}
  \times
  \vcenter{\hbox{\hss\includegraphics[height=15mm]{Pix/xN2B.pdf}\hss}}
  \quad\text{of $N=2$ extended superfields},\\[-2mm]
 \text{then}\quad
 \vcenter{\hbox{\hss\includegraphics[height=15mm]{Pix/N2B22.pdf}\hss}}
  &\Leftrightarrow~
   \big\{(\mathbb{A},\mathbb{B}):
         ~(D_I\mathbb{A})=\varepsilon_{IJ}\delta^{JK}(D_K\mathbb{B})\big\},
\end{align}
where $D_I$ are superderivatives~\cite{r1001,rWB,rBK,rMFTS,rDF}, satisfying $\{D_I,D_J\}=2\delta_{IJ}H$ and $\{D_I,Q_J\}=0=[H,D_I]$. The constrained superfield systems describing the non-maximally extended adinkraic superfields for $N>2$ are conceptually similar, although increasingly more tedious.

Ref.~\cite{r6-1} concluded with conjecturing that there is a superfield of every possible Adinkra topology type. The next section introduces a way to systematically construct Adinkras for which no superfield representation|along the above lines and in traditional superspace|exists in the literature.

\subsection{Projected Supermultiplets}
 \label{s:C}
There also exist adinkraic supermultiplets the Adinkras of which are not merely different ``hangings'' of those in the sequence~(\ref{eA1-5}). The simplest such new Adinkra occurs for $N=4$:
\begin{equation}\setlength{\unitlength}{1mm}
 \vcenter{\hbox{\hss
  \begin{picture}(140,30)(0,-5)
   \put(0,0){\includegraphics[height=22mm]{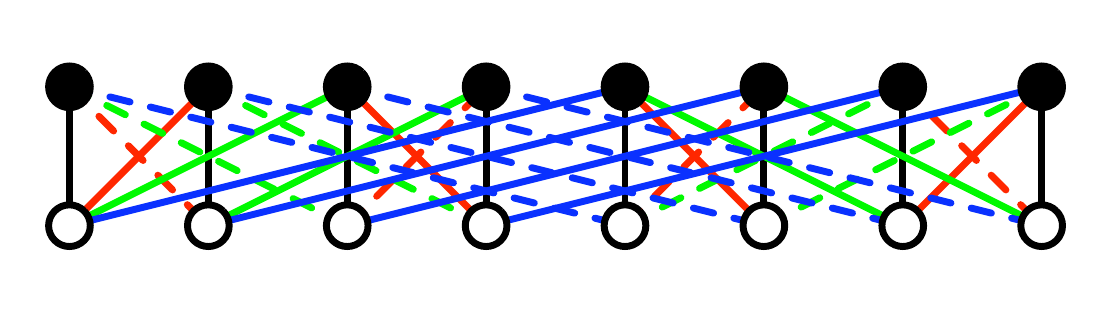}\hss}
    \put(00.5,0){\scriptsize$[0000]$}
    \put(09.5,0){\scriptsize$[0011]$}
    \put(19.5,0){\scriptsize$[0101]$}
    \put(29.5,0){\scriptsize$[0110]$}
    \put(39.5,0){\scriptsize$[1001]$}
    \put(49.5,0){\scriptsize$[1010]$}
    \put(59.5,0){\scriptsize$[1100]$}
    \put(69.5,0){\scriptsize$[1111]$}
    \put(00.5,20){\scriptsize$[0001]$}
    \put(09.5,20){\scriptsize$[0010]$}
    \put(19.5,20){\scriptsize$[0100]$}
    \put(29.5,20){\scriptsize$[0111]$}
    \put(39.5,20){\scriptsize$[1000]$}
    \put(49.5,20){\scriptsize$[1011]$}
    \put(59.5,20){\scriptsize$[1101]$}
    \put(69.5,20){\scriptsize$[1110]$}
   \put(85.5,10){$\longrightarrow$}
   \put(87,12){$\scriptstyle\mathbb{Z}_2$}
   \put(99.5,0){\includegraphics[height=22mm]{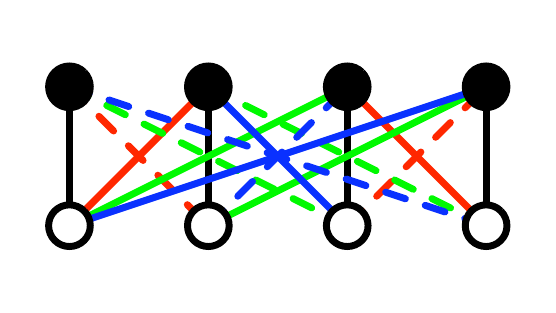}\hss}
    \put(100.5,0){\scriptsize$[0000]$}
    \put(109.5,0){\scriptsize$[0011]$}
    \put(119.5,0){\scriptsize$[0101]$}
    \put(129.5,0){\scriptsize$[0110]$}
    \put(100.5,20){\scriptsize$[0001]$}
    \put(109.5,20){\scriptsize$[0010]$}
    \put(119.5,20){\scriptsize$[0100]$}
    \put(129.5,20){\scriptsize$[0111]$}
  \end{picture}\hss}}
 \label{e2N4As}
\end{equation}
The supermultiplet depicted on the right-hand side is a $\mathbb{Z}_2$-projection of the one on the left-hand side~\cite{r6-3}. The Adinkras, where all the bosons and all the fermions are ``packed'' into two adjacent heights such as the pair~(\ref{e2N4As}), are called {\em\/Valises\/}.

There is an important topological distinction between the two graphs~(\ref{e2N4As}): the left-hand side one decomposes if all edges of any one given color are erased; to decompose the right-hand side Adinkra, one must erase all edges of some {\em\/two\/} colors.
Now, the nodes in both {\em\/Valise\/} Adinkras~(\ref{e2N4As}) can be repositioned at all the combinatorially many various heights to construct the collection of all Adinkras with the two given topology types, as guaranteed by Theorem~5.1 and its Corollary~5.2 in Ref.~\cite{r6-1}. However, the projection map connecting the Valise supermultiplets~(\ref{e2N4As}) may not commute with the vertex-raising operations required to obtain a particular ``hanging'' of either Adinkra, and so only relatively few of the variously ``hanged'' Adinkras will continue to be related by a $\mathbb{Z}_2$-projection, {\em\/i.e.\/}, double cover.

The details of this and analogous projections for $N>4$ are specified in Refs.~\cite{r6-3,r6-3.2}.
 Suffice it here to say that these details of the $\mathbb{Z}_2$-projections such as~(\ref{e2N4As}) are encoded by doubly-even linear binary block codes, $\mathscr{C}$. Therefore, a classification of these codes is necessary for the classification of possible so-projected supermultiplets. The number of such codes grows hyper-exponentially with the number of supersymmetries, $N$, and presents the other source of the tremendous combinatorial complexity of the classification problem; these codes are currently being computed by distributed computation methods that include a supercomputing cluster.

The combination of ({\small\bf1})~the growing number of inequivalent doubly-even linear binary block codes and ({\small\bf2})~the growing complexity of the possible inequivalent ``hangings'' of Adinkras corresponding to any given code is then seen to cause the combinatorial explosion of inequivalent adinkraic supermultiplets:
 Ref.~\cite{r6-3} finds that there exist at least several trillion adinkraic supermultiplets for $N\leq32$. From these, many more can be constructed ``by tensoring and linear algebra.'' This seems much more than what has been indicated in superspace studies so far: besides offering insufficient structure~\cite{rGLP}, the traditional notion of superspace then also seems to offer insufficient {\em\/space\/} to represent all these supermultiplets.

\subsection{Supermultiplets {\em\/vs.\/} Superfields}
The foregoing demonstrates that classifying supermultiplets, while a surprisingly complex problem, did turn out to be more effective than attempting to classify superfields. Even in the case of fixed $N$, there appears to exist no {\em\/a priori\/} exhaustive classifying system for the types of constraints that can be imposed; see for example Ref.~\cite{rHSS}.

In fact, historically, much of the development in the study of supersymmetry and supersymmetric systems did dispense with all reference to superspace, and focused instead on supermultiplets. Views even arose to the effect that the anticommuting coordinates, $\theta$ are merely an inessential bookkeeping artifice. This may well be rooted in the oft-used ``Noether method,'' wherein one starts from a collection of known particles/fields with a known non-supersymmetric action functional, $\mathcal{S}_0$. One then applies the supersymmetry transformations to the action functional, obtaining $Q_I(\mathcal{S}_0)$. If $\mathcal{S}_0$ is not supersymmetric, {\em\/i.e.\/}, if $Q_I(\mathcal{S}_0)\neq\int{\rm d}\tau\,\partial_\tau(\mathcal{K}_I)$, one seeks a {\em\/counterterm\/}, $\mathcal{S}_1$, such that $Q_I(\mathcal{S}_1)$ would provide terms that cancel $Q_I(\mathcal{S}_0)$. If now $Q_I(\mathcal{S}_0{+}\mathcal{S}_1)\neq
\int{\rm d}\tau\,\partial_\tau(\mathcal{K}_I)$, one continues adding counter-terms, aiming to obtain a fully supersymmetric action functional, $\mathcal{S}_0+\mathcal{S}_1+\cdots$. Needless to say, this iterative approach|and in particular the finding of ``correct'' collection of counter-terms|has no formal guarantee and estimate of completion, and is often more of an art-form than science.

In turn, action functionals for superfields are relatively straightforward to compose: An action functional is expressed as an integral over superspace of a Lagrangian super-density, which in turn is a functional expression involving the selected superfields and their superderivatives. Ref.~\cite{r6-4} shows how to use Adinkras to construct proper, physically acceptable kinetic terms for extended superfields of arbitrarily $N$-extended worldline supersymmetry. Still, much remains to be uncovered in this regard.

It behooves us then to explore the correspondence between superspace and supermultiplet methods, as in Refs.~\cite{r6-1,r6-2,r6-7a}. To this end, the seeming mismatch indicated above then motivates a reconsideration of the notion of superspace.

\section{Superspace, by Construction}
 \label{s:SSpace}
Herein, I revisit {\em\/what\/} superspace could be, using only the supersymmetry algebra~(\ref{eSuSy}) and the {\em\/canonical commutation relation\/}~(\ref{eCCR}), and deferring additional assumptions about it for particular applications. It is fortuitous that both ``ingredients'', (\ref{eSuSy}) and~(\ref{eCCR}), are specified in the form of super-commutator brackets. The consequences of these may then be explored simply by examining exhaustively all the graded Jacobi identities\footnote{Herein, the phrase ``Jacobi identities'' will stand in for the much longer locution, ``the consistency of the supersymmetry algebra~(\ref{eSuSy}) with the canonical commutation relation~(\ref{eCCR}),'' unless otherwise specified.} and nothing else.

\subsection{Superpartners of Time}
 \label{s:STime}
The proper time coordinate, $\tau$, is but a particular, real-valued function over the spacetime, $\tau:X\to\mathbb{R}$. In any physical model that exhibits supersymmetry, it makes sense to inquire what are the superpartners of time, $\tau$. That is, as a bosonic function over spacetime, $\tau$ itself must be a component of a supermultiplet, {\em\/i.e.\/}, a representation of supersymmetry, and we then aim to uncover this supermultiplet.

\subsubsection{First Order}
To this end, {\em\/define\/}\footnote{Herein, ``$\equiv$'' is used to mean ``equals identically.''}:
\begin{equation}
  \tau'_I := Q_I(\tau) ~\equiv~ [Q_I,\tau].
 \label{eQt}
\end{equation}
Since Eq.~(\ref{eQt}) is of the form of the left-hand side of~(\ref{eQQB}), this identity and the canonical commutation relation~(\ref{eCCR}) jointly imply:
\begin{equation}
 \{Q^\9_I,\tau'_J\}=\delta_{IJ}\,i\hbar + \{Q^\9_{[I},\tau'_{J]}\}~.
  \label{eXQt'}
\end{equation}
When $I=J$, $\{Q^\9_I,\tau'_J\}$ is nonzero. Therefore, the $\tau'_I$ cannot consistently be set to zero. Furthermore,
\begin{equation}
 [H,\tau'_I]
 =\big[H,[Q_I,\tau]\big]
 \buildrel{\text{(\ref{eBBF})}}\over=-\big[Q_I,[\tau,H]\big]
    -\big[\tau,[H,Q_I]\big]
 =\big[Q_I,i\hbar\big]-\big[\tau,0\big]=0~. \label{eHt'}
\end{equation}
Since $H=i\hbar\partial_\tau$, Eq.~(\ref{eHt'}) means that $\frac{\partial\tau'_I}{\partial\tau}=0$, {\em\/i.e.\/}, $\tau'_I$ and $\tau$ are mutually independent variables.

Perhaps not surprisingly, the foregoing proves:
\begin{proposition}
Consistency of supersymmetry and the canonical commutation relations jointly imply the existence of $\tau'_I$, first-order superpartners to $\tau$. This extends the base spacetime (here, in fact, just time) to a superspace, with coordinates $(\tau|\tau')$ at least.
\end{proposition}
That is, supersymmetry implies, for consistency, that spacetime (here, time) is in fact part of a supermultiplet, identified as superspace.
 This result looks suspiciously like the traditional superspace, $(x|\theta)$, where the $\theta$'s span a vector space dual to $\Span(Q_1,\cdots,Q_N)$. However, in that formalism, the second term on the right-hand side of Eq.~(\ref{eXQt'}) automatically vanishes and $\{Q_I,\theta^J\}=\delta_I{}^J$, for all $I,J=1,\cdots,N$.

\subsubsection{Higher Orders}
Turn to examine the nature of this consistency-implied extension of spacetime: the nature of $\tau'_I$, and the second term in the right-hand side of Eq.~(\ref{eXQt'}):
\begin{equation}
 \tau''_{[IJ]}
  :=\big\{Q^\9_{[I},\tau'_{J]}\big\}
  \equiv\big\{Q_{[I},[Q_{J]},\tau]\big\}
  \equiv Q^\9_{[I}(\tau'_{J]})
  \equiv Q_{[I}Q_{J]}(\tau)~. \label{et''}
\end{equation}
seen to be the second order superpartner of $\tau$.
 As with $\tau'_I$, consider
\begin{equation}
 [Q_I,\tau''_{[JK]}]=[Q_I,\{Q_J,\tau'_K\}]
\end{equation}
by examining the Jacobi identity:
\begin{align}
 0&=\big[Q_I,\{Q_J,\tau'_K\}\big]
   +\big[Q_J,\{\tau'_K,Q_I\}\big]
   +\big[\tau'_K,\{Q_I,Q_J\}\big]~,\nonumber\\
 &=\big[Q_I,\{Q_J,\tau'_K\}\big]
   +\big[Q_J,\{Q_I,\tau'_K\}\big]
   +2\delta_{IJ}\,\big[\tau'_K,H\big]~.
 \label{eJQQt'}
\end{align}
Since the last term vanishes by~(\ref{eHt'}),
\begin{subequations}
 \label{QsQt'}
\begin{gather}
 0=\big[Q_{(I},\{Q_{J)},\tau'_K\}\big]
  =\tfrac12\big([Q^\9_I,\tau''_{[JK]}] + [Q^\9_J,\tau''_{[IK]}]\big)~,\\[-1mm]
\noalign{\noindent or,\vspace{1mm}}
 [Q^\9_I,\tau''_{[JK]}]
  \buildrel{\text{(\ref{QsQt'})}}\over=
 -[Q^\9_J,\tau''_{[IK]}]~.
\end{gather}
\end{subequations}
Used iteratively,  this implies:
\begin{equation}
 [Q^\9_I,\tau''_{[JK]}]=-[Q^\9_I,\tau''_{[KJ]}]
 \buildrel{\text{(\ref{QsQt'})}}\over=
 +[Q^\9_K,\tau''_{[IJ]}]=-[Q^\9_K,\tau''_{[JI]}]
 \buildrel{\text{(\ref{QsQt'})}}\over=
 +[Q^\9_J,\tau''_{[KI]}]=-[Q^\9_J,\tau''_{[IK]}]~,
\end{equation}
and so, in fact,
\begin{align}
 [Q^\9_I,\tau''_{[JK]}]
 &=[Q^\9_{[I},\tau''_{JK]}]
 =\big[Q^\9_{[I},\{Q^\9_J,\tau'_{K]}\}\,\big]~,\\
 &=\big[Q^\9_{[I},\{Q^\9_J,[Q_{K]},\tau]\}\,\big]
  = Q_{[I}Q_JQ_{K]}(\tau)=:\tau'''_{[IJK]}
\end{align}
is the third order superpartner of $\tau$.

The procession $\{\tau,\tau'_I,\tau''_{[IJ]},\tau'''_{[IJK]},\cdots~\}$ continues in this fashion, and it is easy to show that all the $\tau^{\sss[k]}_{[I_1\cdots I_k]}$'s commute with $H$:
\begin{align}
 [H,\tau^{\sss[k]}_{[I_1\cdots I_k]}]
 &=\big[H,[Q_{I_1},\tau^{\sss[k-1]}_{[I_2\cdots I_k]}\}
                -\delta_{k,2}\,\delta_{I_1I_2}\,i\hbar\,\big]
  =\big[H,[Q_{I_1},\tau^{\sss[k-1]}_{[I_2\cdots I_k]}\}\big],\nonumber\\
 &=-\big[Q_{I_1},[\tau^{\sss[k-1]}_{[I_2\cdots I_k]},H]\,\big\}
   +(-1)^k\big[\tau^{\sss[k-1]}_{[I_2\cdots I_k]},\underbrace{[H,Q_{I_1}]}_{=0}\,\big\},
   \nonumber\\
 &=\big[Q_{I_1},[H,\tau^{\sss[k-1]}_{[I_2\cdots I_k]}]\,\big\}
  =\Big[Q_{I_1},\big[Q_{I_2},[H,
         \tau^{\sss[k-2]}_{[I_3\cdots I_k]}]\,\big\}\,\Big\}=\dots,\nonumber\\
 &=\Big[Q_{I_1},\big[Q_{I_2},\cdots\underbrace{[H,
         \tau'_{I_k}]}_{=0}\,\cdots\,\big\}\,\Big\}~=~0~. \label{sHQ-Qt}
\end{align}
Here $[~\,,\>\}$ denotes the super-commutator bracket: an anticommutator if both terms are fermions, and a commutator otherwise.

This also implies that the sequence~(\ref{eSS1}) indeed terminates at the $N^\text{th}$ application of $Q_I$ on $\tau$: By~(\ref{eSuSy}) alone, all $Q$-monomials of degree higher than $N$ must contain at least one copy of $H$, and Eq.~(\ref{sHQ-Qt}) then implies that the action (evaluation) of any such monomial (other than $H$ alone) on $\tau$ will vanish. The (anti-)commutator relations thus derived are summarized in Table~\ref{t1}.
\begin{table}[ht]
  \centering
  \begin{tabular}{r|c@{\quad}c@{\quad}c@{\quad}c@{\quad}c@{\quad}c@{\quad}c@{\quad}c}
    \toprule
 $[~\,,~\}$ & $\tau$ & $\tau'_I$ & $\tau''_{[IJ]}$
          & $\cdots$ & $\tau^{\sss[k]}_{[I_1\cdots I_k]}$
          & $\cdots$ & $\tau^{\sss[N-1]}_{[I_1\cdots I_{N-1}]}$
                     & $\tau^{\sss[N]}_{[I_1\cdots I_N]}$ \\ 
    \midrule
 $Q_L$      & $\tau'_L$ & $\delta_{LI}\,i\hbar+\tau''_{[LI]}$ & $\tau'''_{[LIJ]}$
 & $\cdots$ & $\tau^{\sss[k+1]}_{[LI_1\cdots I_k]}$
 & $\cdots$ & $\tau^{\sss[N]}_{[I_1\cdots I_N]}$ & 0 \\
 $H$        & $i\hbar$ & 0 & 0 & $\cdots$ & 0 & $\cdots$ & 0 & 0 \\
    \bottomrule
  \end{tabular}
  \caption{The (anti-)commutator action of $Q_L$ and $H$ on the objects in the top row.}
  \label{t1}
\end{table}

This produces a complete $Q$-orbit, starting with $\tau$:
\begin{equation}
 \begin{aligned}
 \wQ \,(\tau)
 &:=\{\tau,\tau'_I,\tau''_{[IJ]},\tau'''_{[IJK]},
      \cdots,\tau^{\sss[N]}_{[I_1\cdots I_N]}\}\\
 &~=\{\tau,Q_I(\tau),Q_{[I}Q_{J]}(\tau),Q_{[I}Q_JQ_{K]}(\tau),
      \cdots,Q_{[I_1}\cdots Q_{I_N]}(\tau)\}.
 \label{eSS1}
 \end{aligned}
\end{equation}
This supermultiplet of coordinates thus consists of $2^N$ variables, $2^{N-1}$ bosonic and $2^{N-1}$ fermionic, just as expected of an unconstrained supermultiplet of $N$-extended worldline supersymmetry.

The engineering dimensions of $\tau^{\sss[k]}_{[I_1\cdots I_k]}$, the coordinates of $\wQ (\tau)$, are:
\begin{equation}
  [\tau]=-1,\quad [\tau'_I]=-\tfrac12,\quad [\tau''_{[IJ]}]=0,\quad\dots\quad
  [\tau^{\sss[k]}_{[I_1\cdots I_k]}]=\tfrac{k}2-1,\quad 0\leq k\leq N.
 \label{eEDt}
\end{equation}

Whereas Eq.~(\ref{eXQt'}) implied that $\tau'_I$ cannot consistently be set identically to zero, the analogous Jacobi identity~(\ref{eJQQt'}) implies no such thing for $\tau''_{[IJ]}$, owing to the difference:
\begin{equation}
  [H,\tau] = i\hbar,\qquad\text{\it vs.}\quad
  [H,\tau^{\sss[k]}_{[I_1\cdots I_k]}] = 0,\quad\text{for}\quad k=1,\cdots,N.
\end{equation}
Therefore, the sequence of coordinates
 $(\tau|\tau'_I|\cdots|\tau^{\sss[k]}_{[I_1\cdots I_k]}|\cdots|
                        \tau^{\sss[N]}_{[I_1\cdots I_N]})$
 may be terminated|{\em\/by hand\/}|at any $k>1$; indeed, Table~\ref{t1} implies that setting $\tau^{\sss[k]}_{[I_1\cdots I_k]}=0$ ensures that also
 $\tau^{\sss[\ell]}_{[I_1\cdots I_\ell]}=0$, for $1<k\leq\ell\leq N$.

No further restrictions emerge from the Jacobi identities; see Table~\ref{t:tk}.
\begin{table}[ht]
$$
  \begin{array}{@{} rcc|rcc @{}}
    \toprule{}
    \textbf{Jacobi} & \multicolumn{2}{c|}{\textbf{Consequences}}
  & \textbf{Jacobi} & \multicolumn{2}{c}{\textbf{Consequences}} \\[0mm]
    \midrule{}
 [H,Q_I,\tau] & [H,\tau'_I]=0 & &
 [Q_I,Q_J,\tau] & \boldsymbol{\tau''_{[IJ]}}:=[Q^\9_{[I},\tau'_{J]}]
& \{Q^\9_{(I},\tau'_{J)}\}\neq0\\[1mm]
 [H,Q_I,\tau'_J] & [H,\tau''_{[IJ]}]=0 & &
 [Q_I,Q_J,\tau'_K] & \boldsymbol{\tau'''_{I|[JK]}}:=[Q^\9_I,\tau''_{[JK]}]
& \tau'''_{I|[JK]}=-\tau'''_{J|[IK]}=\tau'''_{[IJK]}\\[1mm]
 [H,Q_I,\tau''_{[JK]}] & [H,\tau'''_{[IJK]}]=0 & &
 [Q_I,Q_J,\tau''_{[KL]}] & \boldsymbol{\tau''''_{I|[JKL]}}:=[Q^\9_I,\tau'''_{[JKL]}]
& \tau''''_{I|[JKL]}=-\tau''''_{J|[IKL]}=\tau''''_{[IJKL]}\\[0mm]
    \bottomrule
  \end{array}
$$
\caption{The systematic list of Jacobi identities using the supersymmetry relations~(\ref{eSuSy}) and the canonical commutation relations~(\ref{eCCR}), and their consequences pertaining to the superspace extension of the worldline, $\mathbb{R}^1$, into the space $\wQ(\tau)$, equipped with a complete supermultiplet of $2^N$ coordinates $\{\tau,\tau'_I,\cdots,\tau^{\sss[N]}_{[I_1\cdots I_N]}\}$.}
 \label{t:tk}
\end{table}

\subsubsection{Comparison with the Traditional Superspace}
It is thus perfectly consistent to set $\tau''_{[IJ]}=0$, whereupon $\wQ (\tau)$ truncates to $\Span(\tau|\tau'_1,\cdots,\tau'_N)$. Furthermore, by defining
\begin{equation}
  \theta^I = \tfrac1{i\hbar}\delta^{IJ}\tau'_J,\qquad
   \text{while}\quad \tau''_{[IJ]}=0~\Rightarrow~
                      \tau^{\sss[k]}_{[I_1\cdots I_k]}=0,~k\geq2,
\end{equation}
the traditional superspace relationship follows:
\begin{equation}
  \{Q_I,\theta^J\} = \tfrac1{i\hbar}\delta^{JK}\,\{Q_I,\tau'_K\}=\delta_I{}^J,\qquad
   \text{if}\quad\tau''_{[IJ]}=0.
 \label{et=q}
\end{equation}
Therefore,
\begin{equation}
  \{\wQ (\tau)\,:~\tau''_{[IJ]}=0\} \simeq \Span(\tau|\theta)
\end{equation}
is indeed the traditional, $(1|N)$-dimensional superspace from the physics literature~\cite{rSSSS1,r1001,rPW,rWB,rBK}, and $\tau''_{[IJ]}$ may be identified as the {\em\/obstruction\/} to $\tau'_I$ being the canonical conjugate to $Q_I$.\bigskip

Evidently, $\wQ (\tau)$ is much bigger than $(\tau|\theta)$. Also, its engineering dimensions-induced $\tfrac12\mathbb{Z}$-grading and associated graded dimension
 $\big(1\big|N\big|\binom{N}2\big|\cdots\big|\binom{N}{N-1}\big|1\big)$ {\em\/mirrors\/} the superfield expansion~(\ref{eSFX}). Note: $\tau^{\sss[k]}_{[I_1\cdots I_k]}$ cannot be identified with the antisymmetric product
 $(\wedge^k\tau')_{[I_1\cdots I_k]}=\tau'_{[I_1}\cdots\tau'_{I_k]}$, since
\begin{alignat}{5}
  [\tau^{\sss[k]}_{[I_1\cdots I_k]}]&=\tfrac{k}2{-}1,&
   \qquad&\text{whereas}&\qquad
  [\tau'_{[I_1}\cdots\tau'_{I_k]}]&=k(-\tfrac12)=-\tfrac{k}2.\\[-1mm]
\noalign{\noindent They agree only for $k=1$.
It follows that the $2^N$ coordinates of $\wQ (\tau)$ are all|{\em\/a priori\/}|algebraically independent and not identically zero. Also, the engineering dimensions of $\wedge^k\theta$ and $\tau^{\sss[k]}$ vary oppositely:\vspace{1mm}}
 [\wedge^{k+1}\theta]&<[\wedge^k\theta],&
   \qquad&\text{whereas}&\qquad
 [\tau^{\sss[k+1]}]&>[\tau^{\sss[k]}],
 \label{eOpED}
\end{alignat}
whereupon the engineering dimension-based grading in $\wQ (\tau)$ quite literally {\em\/mirrors\/} that of $\mathcal{O}_X[\wedge\theta]$:
\begin{equation}
 \begin{array}{ccrclrclc}
 \textbf{Eng.~dim.}
 &\textbf{dim.}
 &\multicolumn{3}{c}{\boldsymbol{\wQ(\tau)}\textbf{-basis}^*}
 &\multicolumn{3}{c}{\qquad\boldsymbol{\wedge\theta}\textbf{-basis}}
 &\textbf{dim.}\\
 \toprule
 \vrule width0pt height3ex
 (\frac{N}2{-}1) & 1
 &\tau^{\sss[N]}_{[I_1\cdots I_N]} &\in &\wedge^NQ\,(\tau) & &\text{|} \\
  ~~\vdots &\vdots& &\vdots& & &\vdots \\
 +\frac12 &\binom{N}{3} &\tau'''_{[IJK]} &\in &\wedge^3Q\,(\tau) & &\text{|}\\[1mm]
 0 &\binom{N}2 &\tau''_{[IJ]} &\in &\wedge^2Q\,(\tau) &1 &= &\Ione &1 \\
 -\frac12 &N &\tau'_I &\in &Q(\tau) &\theta^I &\in &\theta &N \\
 -1 &1 &\tau &= &\tau &\theta^{[I}\theta^{J]} &\in &\wedge^2\theta &\binom{N}2 \\[1mm]
 -\frac32 & & &\text{|}& &\theta^{[I}\theta^J\theta^{K]} &\in &\wedge^3\theta
  &\binom{N}3 \\
  ~~\vdots & & &\vdots& & &\vdots & &\vdots\\
 -\frac{N}2 & 0 & &\text{|} & & \theta^{[I_1}\cdots\theta^{I_N]}&\in&\wedge^N\theta
  & 1 \\
 \bottomrule
 \multicolumn{9}{l}{{}^*\text{\small For brevity, only the generators of the $\wQ(\tau)$-basis are shown, not their products.}}
 \end{array}
 \label{eMQt}
\end{equation}
and $\mathcal{O}_X[\wedge\theta]$ denotes formal linear combination of the elements of $\wedge\theta$, with coefficients that are functions over $X$|precisely the familiar $\theta$-expansion of a superfield over the traditional superspace. Since
\begin{equation}
  \wedge\big(Q(\tau)\big) \simeq \wedge\theta,
\end{equation}
all the ``higher'' superpartners of time, $\tau''_{[IJ]},\tau'''_{[IJK]},\cdots$ then extend the traditional superspace.

It should be noted that the $\mathbb{Z}$-grading introduced by the filtration and quotient-grading~(\ref{eGr}) agrees with with the double of the engineering dimension-induced $\tfrac12\mathbb{Z}$-grading only over the traditional superspace, parametrized by $(\tau|\theta^I)$|when $\tau^{\sss[k]}_{[I_1\cdots I_k]}=0$ for $k\geq2$. This is because only over this, first-order truncation of $\wQ(\tau)$ do all odd (fermionic) coordinates $\theta^I:=\delta^{IJ}\tau'_J/i\hbar$ have the same engineering dimension, $[\theta^I]=-\tfrac12$.
 One defines $\mathcal{F}_M:=\Gr_1\mathcal{O}_M=J_M/J^2_M$. Given a base $\dim(M)=(n|N)$, $\mathcal{F}_M$ is a locally free sheaf of rank $0|N$, identifiable with the one spanned by the $\theta^I$'s over $(\tau|\theta)$. One then finds that
 $\Gr_\ell\mathcal{O}_M\simeq\mathcal{O}_M[\wedge^\ell\mathcal{F}]$, {\em\/i.e.\/}, formal linear combinations of elements of $\wedge^\ell\mathcal{F}$ with ordinary functions over $M$ as coefficients.
 In the traditional physics notation, elements of $\Gr_\ell\mathcal{O}_M$ are then the $\wedge^\ell\theta$-coefficients, {\em\/i.e.\/}, the component fields which multiply $\theta^{[I_1}\cdots\theta^{I_\ell]}$ in the expansion~(\ref{eSFX}), and where $[\wedge^\ell\theta]=-\frac{\ell}2$.

However, this agreement no longer holds when comparing $\Gr_\ell\mathcal{O}_M$ with degree-$\ell$ homogeneous elements of $\wQ (\tau)$. In stark contradistinction to $(\tau|\theta)$, there are many more|and much more varied|general monomials of a generic, fixed engineering degree in $\wQ (\tau)$:
\begin{equation}
  \mathfrak{m}:=\prod_{k=1}^N(\Pi^{p_k}\tau^{\sss[k]}),\qquad
   [\mathfrak{m}]
   =\sum_{k=1}^N p_k\,[\tau^{\sss[k]}_{[I_1\cdots I_k]}]
   =\sum_{k=1}^N p_k\,\big(\tfrac{k}{2}{-}1\big).
 \label{etMon}
\end{equation}
In particular, note that:
\begin{itemize}\itemsep=-3pt
 \item The power $p_2$, of $\tau''_{[IJ]}$'s in $\mathfrak{m}$, is
       {\em\/not restricted\/} by the condition 
       $[\mathfrak{m}]=-\frac\ell2$, since $[\tau''_{[IJ]}]=0$.
 \item Unlike $(\tau|\theta)$, $\wQ (\tau)$ also has coordinates with non-negative
       engineering dimension when $N\geq2$.
 \item Although the $\tau^{\sss[k]}_{[I_1,\cdots I_k]}$'s are bosonic when $k$
       is even, and fermionic when $k$ is odd, it has not yet been determined which
       of these variables are super-commuting, and which nilpotent. Once determined,
       this in turn determines the type of product $\Pi^{p_k}$ in~(\ref{etMon})
       to be $\wedge^{p_k}$, $\Sym^{p_k}$ or perhaps just $\otimes^{p_k}$, with
       perhaps a nilpotence-induced termination limit.
\end{itemize}
This last aspect is what we explore next.

\subsection{A Telescoping Deformation Structure}
 \label{s:SMod}
Since $\tau^{\sss[k]}_{[I_1\cdots I_k]}=Q_{[I_1}\cdots Q_{I_k]}(\tau)$ is the $k^\text{th}$ $Q$-transformation of $\tau$, it is bosonic for even $k$, and fermionic for odd $k$. One would therefore expect these variables to be mutually super-commuting:
\begin{equation}
 \zeta_1\,\zeta_2 - (-1)^{|\zeta_1||\zeta_2|}\,\zeta_2\,\zeta_1 = 0,
\end{equation}
where $\zeta_1,\zeta_2$ are some two {\itshape homogeneous\/} elements of $\wQ(\tau)$, and $|\zeta|$ is the degree of homogeneity, such that
 $|\tau^{\sss[k]}_{[I_1\cdots I_k]}|=k$, since $|H|=0=|\tau|$ and $|Q_I|=1$. As above, it behooves to determine what in this respect is implied by the Jacobi identities.

\subsubsection{Super-Commutivity Obstructions}
For future convenience, write
\begin{equation}
 \eta:=[\tau,\tau], \label{ee0}
\end{equation}
and record that $\eta=0$, albeit trivially.

Next, one would {\em\/expect\/} $\tau'_I$ to commute with $\tau$, so we define:
\begin{equation}
 \eta'_I~:=~[\tau,\tau'_I]\equiv[\tau,Q_I(\tau)]:\qquad
  \text{obstruction to }[\tau,\tau'_I]=0. \label{eeta1}
\end{equation}
A direct evaluation of $\eta'_I$ by means of the definition~(\ref{eQt}) and the Jacobi identity~(\ref{eBBF}) starting with ``$\big[\tau,[Q,\tau]\big]$'' fails, as this identity is trivially satisfied:
\begin{equation}
 0 =\big[\tau,[Q_I,\tau]\,\big]
   +\big[Q_I,\underbrace{[\tau,\tau]}_{=0}\,\big]
   +\big[\tau,[\tau,Q_I]\,\big]
  ~=~\big[\tau,[Q_I,\tau]\,\big]
   -\big[\tau,[Q_I,\tau]\,\big]~\equiv~0~.
\end{equation}
On the other hand, the Jacobi identity starting with $[H,\eta'_I]=\big[H,[\tau,\tau'_I]\,\big]$ does provide information:
\begin{align}
 0&=\big[H,[\tau,\tau'_I]\,\big]
   +\big[\tau,[\tau'_I,H]\,\big]
   +\big[\tau'_I,[H,\tau]\,\big]
  ~=~\big[H,[\tau,\tau'_I]\,\big]
   +\big[\tau,0\,\big]
   +\big[\tau'_I,i\hbar\,\big]~,\nonumber\\
 &\Rightarrow
 [H,\eta'_I]=0.\label{eHtt'}
\end{align}
That is, the Jacobi identities imply only that the $\eta'_I$'s are $\tau$-independent.

Since the $\tau'_I=Q_I(\tau)$ are fermionic, we {\em\/expect\/} them to anti-commute. To this end, we use the first version of the Jacobi identity~(\ref{eFFB}) to express:
\begin{equation}
 \{\tau'_I,\tau'_J\}
  =\{\tau'_I,[Q_I,\tau]\}
  \buildrel{\text{(\ref{eFFB})}}\over=
   \{Q_J,[\tau,\tau'_I]\} - [\tau,\{\tau'_I,Q_J\}]
 ~=~\{Q_J,\eta'_I\} - [\tau,\{\tau'_I,Q_J\}].
\end{equation}
Then, projecting on the $I\leftrightarrow J$ symmetric and antisymmetric part, respectively:
\begin{subequations}
\begin{alignat}{3}
 \{\tau'_I,\tau'_J\}&\equiv&\{\tau'_{(I},\tau'_{J)}\}
 &=\{Q^\9_{(I},\eta'_{J)}\}-\underbrace{[\tau,i\delta_{IJ}\hbar]}_{=0}
 ~=~Q^\9_{(I}(\eta'_{J)});\\
 0&\equiv&\{\tau'_{[I},\tau'_{J]}\}
 &=\{Q^\9_{[I},\eta'_{J]}\}-[\tau,\tau''_{[IJ]}]
 ~=~Q^\9_{[I}(\eta'_{J]})-[\tau,\tau''_{[IJ]}].
\end{alignat}
\end{subequations}
Neither of these quantities is any further determined by the Jacobi identities, and we {\em\/define\/}:
\begin{subequations}
 \label{eeta2}
\begin{alignat}{5}
 \eta''_{(IJ)}&:=Q^\9_{(I}(\eta'_{J)})&&~=~\{\tau'_I,\tau'_J\},&\qquad
  \text{obstruction to }\{\tau'_I,\tau'_J\}&=0,\\
 \eta''_{[IJ]}&:=Q^\9_{[I}(\eta'_{J]})&&~=~[\tau,\tau''_{[IJ]}],&\qquad
  \text{obstruction to }[\tau,\tau''_{[IJ]}]&=0.
\end{alignat}
\end{subequations}
Next, we find that
\begin{subequations}
 \label{eQe2}
\begin{align}
 Q^\9_{[I}(\eta''_{JK]})
 &=[\tau,\tau'''_{[IJK]}]+[\tau'_{[I},\tau''_{JK]}],\qquad\text{and}\\
 \tfrac13\big[Q^\9_K(\eta''_{(JI)})-Q^\9_J(\eta''_{(KI)})\big]
 &=[\tau'_I,\tau''_{[JK]}]-[\tau'_{[I},\tau''_{JK]}]
\end{align}
\end{subequations}
are similarly undetermined by the Jacobi identities. While the first array of variables is totally antisymmetric, the second one corresponds to the mixed representation of the permutation group, depicted by the Young tableau $[2,1,0,\cdots]:=\raisebox{-1pt}{\includegraphics[height=2.3ex]{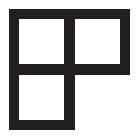}}$, where in ``$[n_1,n_2,\cdots,n_N]$,'' $n_r$ denotes the number boxes in the $r^\text{th}$ row from the top; $n_r\geq n_{r+1}$. Notice, however, that the two arrays of third-order $\eta$'s, $\eta'''_{[IJK]}$ and $\eta'''_{[KJ]I}$ do not completely describe the super-commutativity obstructions amongst $\{\tau,\tau',\tau'',\tau'''\}$: to that end we define:
\begin{alignat}{3}
 \eta^{\sss[3]}_{[IJK]}&:=[\tau,\tau'''_{[IJK]}],&\qquad
  \text{obstruction to }&[\tau,\tau'''_{[IJK]}]=0;\\
 \eta^{\sss[1,2]}_{[IJK]}&:=[\tau'_{[I},\tau''_{JK]}],&\qquad
  \text{obstruction to }&[\tau'_{[I},\tau''_{JK]}]=0;\\
 \eta^{\sss[2]1}_{[KJ]I}&:=[\tau'_I,\tau''_{[JK]}]-[\tau'_{[I},\tau''_{JK]}]&\qquad
  \text{obstruction to }&\pi_{_{\includegraphics[width=2mm]{Pix/YT21.pdf}}}
                          \big([\tau'_I,\tau''_{[JK]}]\big)=0,
\end{alignat}
where ``$\pi_R$'' denotes the projection to the representation $R$ of the permutation group.

Continuing in this vein introduces an array of new variables that grows combinatorially with $N$ and may be arranged into a triangular scheme:
\begin{equation}
  \begin{array}{@{} rl @{}}
    \toprule
 \boldsymbol{[\>\cdot\>]} & \textbf{Super-Commutivity Obstructions} \\ 
    \midrule
        -2 & \eta:=[\tau,\tau],\\
 -\tfrac32 & \eta'_I:=[\tau,\tau'_I] \\ 
        -1 & \eta''_{[IJ]}:=[\tau,\tau''_{[IJ]}],\quad
             \eta''_{(IJ)}:=\{\tau'_I,\tau'_J\}, \\ 
 -\tfrac12 & \eta'''_{[IJK]}:=[\tau,\tau'''_{[IJK]}],\quad
             \eta'''_{I|[JK]}:=[\tau'_I,\tau''_{[JK]}] \\ 
         0 & \eta''''_{[IJKL]}:=[\tau,\tau''''_{[IJKL]}],\quad
             \eta''''_{I|[JKL]}:=\{\tau'_I,\tau'''_{[JKL]}\},\quad
             \eta''''_{[IJ]|[KL]}:=[\tau''_{[IJ]},\tau''_{[KL]}] \\
    \vdots & \vdots \\
 \tfrac{N}2{-}2 & \eta^{\sss[N]}_{[I_1\cdots I_N]}
             :=[\tau,\tau^{\sss[N]}_{[I_1\cdots I_N]}\},\quad
              \eta^{\sss[1,N-1]}_{I_1|[I_2\cdots I_N]}
              :=\big[\tau'_{I_1},\tau^{\sss[N-1]}_{[I_2\cdots I_N]}\big\},\quad
              \cdots\quad
              \eta^{\sss[N/2,N/2]}_{[I_1\cdots I_N]}
              :=\big[\tau^{\sss\lfloor N/2\rfloor}_{[\cdots]},
               \tau^{\sss\lceil N/2\rceil}_{[\cdots]}\big\}\\
    \vdots & \vdots \\
 N{-}\tfrac52 & \eta^{\sss[N-1,N]}_{[\cdots]|[\cdots]}
    :=[\tau^{\sss N-1}_{[\cdots]},\tau^{\sss N}_{[\cdots]}\}\\[2mm]
    N{-}2  & \eta^{\sss[N,N]}_{[\cdots]|[\cdots]}
    :=[\tau^{\sss N}_{[\cdots]},\tau^{\sss N}_{[\cdots]}\},\quad
    \text{(nontrivial for odd $N$)}\\
    \bottomrule
 \multicolumn{2}{l}{\text{\footnotesize For clarity, the entries have not been projected to irreducible representations of the permutation group.}}
  \end{array}
 \label{eSO}
\end{equation}
These obstructions are, quite clearly, functions over $\wQ(\tau)$, but some subset of them may be set equal to certain constant values. For example, setting
\begin{equation}
   \eta''_{(12)}:=\{\tau'_1,\tau'_2\} ~\mapsto~0
\end{equation}
makes $\tau'_1$ and $\tau'_2$ anticommute, but says nothing of the other $\tau'_I$'s. In this way, the space of assignable values of all the obstructions~(\ref{eSO}) forms a parameter space for the super-commutativity of $\wQ(\tau)$. The cases where some of the super-commutators~(\ref{eSO}) remain free functions of their two arguments may be included by redefining the $\eta$'s in a discontinuous way:
\begin{subequations}
 \label{eDisC}
\begin{align}
  \eta''_{(IJ)}:=\{\tau'_I,\tau'_J\}~\mapsto~
  &(1-h''_{(IJ)})\big(\eta''_{(IJ)}:=\{\tau'_I,\tau'_J\}\big)
   =(1-h''_{(IJ)})y''_{(IJ)}\\[1mm]
  &\Rightarrow~
  \bigg\{\begin{array}{rl@{~\text{if}\quad}l}
           \eta''_{(IJ)}:=\{\tau'_I,\tau'_J\}&=y''_{(IJ)}, & h''_{(IJ)}\neq1,\\[1mm]
           \eta''_{(IJ)}:=\{\tau'_I,\tau'_J\}&\text{is free}, & h''_{(IJ)}=1.
         \end{array}
\end{align}
\end{subequations}
Denote by $Y_{[\cdot,\cdot\}}$ the space of assignable values paired with a discontinuous interpolating variable each, such as $(y''_{(IJ)},h''_{(IJ)})$. There is abundant redundancy in this parametrization: for example, $\eta''_{(12)}$ and $\eta''_{(13)}$ are clearly related by the reparametrization $\tau'_2\leftrightarrow\tau'_3$. Modulo such identifications, $Y_{[\cdot,\cdot\}}$ provides an effective parameter space for the possible choices involving the super-commutativity obstructions~(\ref{eSO}).

\subsubsection{Nilpotence and Higher Order Obstructions}
At the origin of $Y_{[\cdot,\cdot\}}$, where all super-commutativity obstructions~(\ref{eSO}) are set to vanish, the coordinates of $\wQ(\tau)$ are super-commutative. In particular, $\tau^{\sss[k]}_{[I_1\cdots I_k]}$ are commuting variables for even $k$. However, the commuting coordinates $\tau^{\sss[k]}_{[I_1\cdots I_k]}$ may or may not be nilpotent: that represents an additional choice.

This makes $\wQ(\tau)$ completely unlike the traditional superspace, where all non-spacetime coordinates are anticommuting, necessarily nilpotent, and so can generate but a finite exterior algebra and admit only polynomials of finite degree. By contrast, $\wQ(\tau)$ admits rather arbitrary functions of some of its non-spacetime coordinates.

Furthermore, the obstructions to nilpotence, such as
\begin{equation}
  (\tau''_{[IJ]})^2,\qquad (\tau''''_{[IJKL]})^2,\quad \textit{etc.},
 \label{eNt}
\end{equation}
cannot possibly be determined from the Jacobi identities, since these identities never involve simple squares of variables.

Just as the obstructions to super-commutativity in $\wQ(\tau)$~(\ref{eSO}), the obstructions to its nilpotence~(\ref{eNt}) are quadratic in the coordinates of $\wQ(\tau)$. Define then a discontinuous parametrization of the nilpotence obstructions akin to~(\ref{eDisC}), and consider the space, $Y$, of pairs---each consisting of an assignable value and a discontinuous interpolation variable---for each of these two types of obstructions jointly.

Since particular choices of values for the various $(y,h)$'s, \textit{i.e.}, various points of $Y$ specify differing super-commutativity structures for $\wQ(\tau)$, it is natural to fiber these variably super-commuting ``copies'' of $\wQ(\tau)$ over $Y$. Thus, the space of super-commutativity obstructions of $\wQ(\tau)$ may be regarded as a deformation space for its super-commuting structure, and the \textit{total space} of this fibration then defines an even bigger extension of of superspace.

Manifestly, one may as reasonably inquire about super-commutativity in the total space of this fibration of $\wQ(\tau)$ over $Y$. This then introduces new pairs of assignable values and corresponding discontinuous interpolating variables, which may in turn be regarded as spanning a deformation space for the fibration of $\wQ(\tau)$ over $Y$. Call this space $Z$; fibrations of $\wQ(\tau)$ over $Y$ may then themselves be fibered over $Z$, and the total space of this fibration over $Z$ may in turn be regarded as an even bigger extension to superspace.

This iterative, telescoping fibration clearly never need stop.

\subsection{Nontrivial Superspace Geometry}
 \label{s:STop}
Traditional superspace is in many ways regarded as lackluster. For one, the space spanned by anticommuting and nilpotent coordinates $\theta^I$ does not seem to offer much opportunity for non-trivial geometry or topology. In part, this has been one of the reasons for easy dismissal of superspace as ``just another bookkeeping device''.
 Even in the traditional superspace, however, there does exist a possibility of non-trivial structure that has not been employed so far. 

The recent work of Refs.~\cite{r6-1,r6-3,r6-3.2} uncovers a combinatorial plethora of representations of $N$-extended worldline supersymmetry without central charges. This classification program focuses on {\em\/adinkraic\/} representations (see Section~\ref{s:A}), for which a notion of {\em\/chromotopology\/} is defined. It turns out that chromotopologies available for adinkraic supermultiplets must be $(\mathbb{Z}_2)^k$-quotients of $N$-cubes, with the quotient actions classified by doubly even binary linear block codes, $\mathscr{C}$.

These quotients may be described as systems of operatorial constraints of the form
\begin{equation}
 Q_1Q_2 - Q_3Q_4 = 0,\qquad
 Q_1Q_3 + Q_2Q_4 = 0,\qquad
 Q_1Q_4 - Q_2Q_3 = 0,
 \label{e:D4}
\end{equation}
imposed on the supersymmetry charges when acting on a representation with the corresponding quotient chromotopology. The left-hand side expressions of the constraints~(\ref{e:D4}) generate an ideal, $\mathscr{I}_{D_4}$, in the universal enveloping algebra generated by the supersymmetry algebra~(\ref{eSuSy}). In particular, applying $Q_I$ on the left\footnote{The ideal generated by~(\ref{e:D4}) is thus a right ideal: elements of the ideal multiply $Q$-polynomials only from the right, {\em\/i.e.\/} $Q$-polynomials multiply elements of the ideal only from the left.} of the constraints, we obtain
\begin{gather}
 H\,Q_2 - Q_1Q_3Q_4 = 0,\quad
 H\,Q_1 + Q_2Q_3Q_4 = 0,\quad
 Q_1Q_2Q_3 - H\,Q_4 = 0,\quad  
 Q_1Q_2Q_4 + H\,Q_3 = 0,
\end{gather}
and from this
\begin{equation}
 H^2 + Q_1Q_2Q_3Q_4 = 0. \label{e:D4P}
\end{equation}
The correspondence to doubly even binary linear block codes is read off from this last equation~(\ref{e:D4P}) by interpreting the $N$-tuple of exponents, $(1,1,1,1,0,\cdots)$, of the $Q_I$'s in~(\ref{e:D4P}) as a binary codeword, $111100\cdots$. The constraints~(\ref{e:D4})--(\ref{e:D4P}) impose a projection on the universal enveloping algebra of the supersymmetry algebra~(\ref{eSuSy}), and thereby define a projected representation. The superspace for the supersymmetry algebra subject to constraints such as~(\ref{e:D4})--(\ref{e:D4P}) then must have a geometry that is consistent with these constraints, and this induces a less than trivial geometry on the superspace.

In the traditional superspace, the fermionic coordinates $\theta^I$ are strictly dual to the supercharges $Q_I$, being a map from the space coordinatized by the $Q_I$'s to $\mathbb{R}$. Therefore, the $\mathscr{C}$-encoded constraints on the supercharges~(\ref{e:D4}) have a dual effect on the  $\theta^I$'s, thus providing even this traditional superspace with a less than trivial algebro-geometric structure.

Manifestly, the $\mathscr{C}$-encoded constraints~(\ref{e:D4}) have a similarly dual induced effect on all of $\wQ(\tau)$, as well as on $Y,Z,\dots$ For example, the operatorial constraints~(\ref{e:D4}) imply that:
\begin{subequations}
 \label{eSD2}
\begin{gather}
 \tau''_{12}=\tau''_{34},\quad
 \tau''_{13}=-\tau''_{24},\quad
 \tau''_{14}=\tau''_{23},\\
 \tau'''_{123}=[H,\tau'_4]=0,\quad
 \tau'''_{124}=-[H,\tau'_3]=0,\quad
 \tau'''_{134}=[H,\tau'_2]=0,\quad
 \tau'''_{234}=-[H,\tau'_1]=0,\\
 \tau''''_{1234}=[H,\tau]=i\hbar=\textit{const}.\neq0.
\end{gather}
\end{subequations}
This halves the list of independently variable coordinates of the $N=4$ superspace $\wQ(\tau)$:
\begin{equation}
  (\tau|\tau'_I|\tau''_{[IJ]}|\tau'''_{[IJK]}|\tau''''_{1234})
  \quad\longrightarrow\quad
  (\tau|\tau'_I|\tau''_{[IJ]}),\quad\tau''_{[IJ]}
   =\tfrac12\varepsilon_{IJ}{}^{KL}\tau''_{[KL]},
 \label{e:D4SS}
\end{equation}
where $\varepsilon_{IJKL}$ is the $N=4$ totally antisymmetric Levi-Civita symbol.

The operatorial constraints~(\ref{e:D4})--(\ref{e:D4P}) in fact describe the $Q_I$-action on a supermultiplet closely related to the worldline restriction of a well-known representation of simple ($\mathcal{N}=1$) supersymmetry in 4-dimensional spacetime: the chiral supermultiplet, which is one of the best-known examples in superspace formulation! It might then come as a surprise that the less than trivial consequences of the operatorial constraints~(\ref{e:D4})--(\ref{e:D4P}) on superspace have never been detected.

However, note that traditional superspace is embedded in $\wQ(\tau)$ as the $\tau''_{[IJ]}=0$ sub-superspace. And, in this sub-superspace, the operatorial constraints~(\ref{e:D4})--(\ref{e:D4P}) have {\em\/no effect\/}.

Recall that the Jacobi identities---consistency of the supersymmetry algebra~(\ref{eSuSy}) together with the canonical commutation relations~(\ref{eCCR})---do not imply the $\tau''_{[IJ]}$ to be either commuting or nilpotent. That is, the quantities
\begin{equation}
 \eta''''_{[IJ]|[KL]}:=[\tau''_{[IJ]},\tau''_{[KL]}],\qquad
 \eta''''_{[IJ]}:=(\tau''_{[IJ]})^2
\end{equation}
are undetermined, apart from $\eta''''_{[IJ]|[KL]}=-\eta''''_{[KL]|[IJ]}$ by definition. The constraints~(\ref{eSD2}) then embed this possibly non-commutative $D_4$-superspace into $\wQ(\tau)$ as a linear sub-superspace of half total dimension. If we further assume that the $\tau''_{12},\tau''_{13},\tau''_{23}$ are ``ordinary'' commuting variables:
\begin{equation}
 \text{assume }:\quad\eta''''_{[IJ]|[KL]}:=[\tau''_{[IJ]},\tau''_{[KL]}]~=~0,
\end{equation}
and that $\eta''''_{[IJ]}:=(\tau''_{[IJ]})^2$ is not constrained to a particular value, the three coordinates $\tau''_{12},\tau''_{13},\tau''_{23}$ may well be ``added'' to the worldline as coordinates of ``ordinary'' spacetime. Being that $[\tau''_{[IJ]}]=0$, these coordinates have no engineering dimension and are akin to hyperbolic angles, {\em\/i.e.\/}, ratios of ordinary length-coordinates.

Power-expansions in $\tau'_I$ of functions of $(\tau|\tau'_I|\tau''_{[IJ]})$ terminate as usual, owing to the now chosen anti-commutativity, $\{\tau'_I,\tau'_J\}=0$, providing the usual superspace expansion. However, power-expansion over $\tau''_{12},\tau''_{13},\tau''_{23}$ does not terminate unless we further assume that all $\eta''''_{[IJ]}=0$, so that the $\tau''_{12},\tau''_{13},\tau''_{23}$ are in fact commuting but nilpotent. In that case, power-expansion over $\tau''_{12},\tau''_{13},\tau''_{23}$ is effectively an expansion in the monomials
\begin{subequations}\vspace{-3mm}
\begin{gather}
  \tau''_{12},\quad \tau''_{13},\quad \tau''_{23},\\
  \tau''_{12}\tau''_{13},\quad \tau''_{12}\tau''_{23},\quad \tau''_{13}\tau''_{23},\\
  \tau''_{12}\tau''_{13}\tau''_{23},
\end{gather}
\end{subequations}
where the products are symmetric, however. Effectively, this adds seven additional component fields obtained in the $\tau''_{[IJ]}$-expansion to every component field over the traditional $N=4$ superspace, $(\tau|\tau'_1,\tau'_2,\tau'_3,\tau'_4)$.

Together with the novelty of the classification efforts based on code-encoded projections using relations of the type~(\ref{e:D4})--(\ref{e:D4P}) in Refs.~\cite{r6-1,r6-3,r6-3.2}, this explains why no semblance of less than trivial geometry has so far been bestowed upon superspace.

\subsection{Higher-Dimensional Spacetime}
 \label{s:Higher}
The supersymmetry super-commutation relations may now be written:
\begin{subequations}
 \label{eSuSyH}
 \begin{alignat}{3}
 \{Q_I,Q_J\} &=2\,\Gamma^j_{IJ}\,P_j~,&\qquad
  \Gamma^i_{IJ}&=\Gamma^i_{JI}~,\quad\forall\, i~,\label{eQQP}\\[1mm]
   [Q_I,P_j] &=0~,&\qquad
   [P_i,P_j] &=0~, \label{eQPP}
\end{alignat}
\end{subequations}
where the Clebsch-Gordan-like coefficients $\Gamma^i_{IJ}$ are generalized Dirac gamma matrices, encoding the relation between the chosen bases for $\Span(Q_1,\cdots,Q_N)$ and $\Span(P_0,\cdots,P_{n-1})$ in the relation
\begin{equation}
 \Span(P_0,P_1,\cdots,P_{n-1}) \subset
  \Sym\nolimits^2\Span(Q_1,\cdots,Q_N)~.
\end{equation}
For each $i$, the matrix $\Gamma^i_{IJ}$ must be invertible. Also, the matrices $[\hbox{I\kern-.18em$\Gamma$}^i]_{IJ}=\Gamma^i_{IJ}$ satisfy an appropriate Clifford algebra and we may furthermore {\em define\/} the matrix-inverse of these generalized Dirac matrices by the relation:
\begin{equation}
 \Gamma_i{}^{IJ}~:\quad
 \Gamma_i^{IJ}\,\Gamma^j{}_{JK}=\delta_i{}^j\,\delta^I{}_K+\eta_{ik}\,\Gamma^{[kj]\,I}{}_K~,
 \label{eGG1}
\end{equation}
where $\eta_{ik}$ is the (preferred) metric on $\Span(P_1,\cdots,P_d)$, and $\Gamma^{[ik]\,I}{}_K$ generate the Lorentz group action on $\Span(Q_1,\cdots,Q_N)$: $Q_K\mapsto\tfrac12\ell_{[ik]}\Gamma^{[ik]\,I}{}_K\,Q_I$.

Reality and other properties of $Q_I$ and $\Gamma^i_{IJ}$, as well as the use of $\gamma^0$ for the Dirac-conjugate of $Q_I$ will depend on the spacetime dimension, $d$, signature and perhaps additional choices. While important in concrete applications, these details are not relevant for our present analysis and would needlessly complicate our ``generic spacetime'' notation.

As before, one considers all the Jacobi identities obtained from the supersymmetry algebra~(\ref{eSuSyH}) and the canonical commutation relations:
\begin{equation}
  [P_j,x^k]=i\hbar\,\delta_j{}^k. \label{eCCRH}
\end{equation}
As before, this implies that the immediate superpartners
\begin{equation}
  \xi'_I{}^i:=[Q_I,x^i]=Q_I(x^i)
 \label{eX'}
\end{equation}
of spacetime coordinates satisfy
\begin{align}
 \{Q^\9_I\,,\,\xi'_J{}^j\}&=\xi''_{[IJ]}{}^j+i\hbar\,\Gamma^j_{IJ}~, \label{eQX'}\\[2mm]
 \text{where}\quad
 \xi''_{[IJ]}{}^j&:=\{Q^\9_{[I},\xi'_{J]}{}^j\}
  =\big\{Q_{[I},[Q_{J]},x^j]\big\}
  =Q^\9_{[I}(\xi'_{J]}{}^j)
  =Q_{[I}Q_{J]}(x^j)~. \label{eX''}
\end{align}
As before, it would be inconsistent to set $\xi'=0$, but setting $\xi''=0$ {\em\/is\/} consistent with the supersymmetry algebra~(\ref{eSuSyH}) and the canonical commutation relations~(\ref{eCCRH}). Proceeding as before, one obtains the superspace $\wQ(X)$, coordinatized by:
\begin{equation}
  x^i,\quad \xi'_I{}^i,\quad \xi''_{[IJ]}{}^i,\quad \cdots\quad
  \xi^{\sss[N]}_{[I_1\cdots I_N]}{}^i,
\end{equation}
forming a $\big(n\big|n\,N\big|n\,\binom{N}2\big|\cdots\big|n\,\binom{N}{N}\big)$-dimensional supermultiplet of coordinates. Traditional superspace, $(x^i|\theta^I)$, is the linear subspace of $\wQ(X)$, obtained by setting:
\begin{alignat}{3}
 \xi''_{[IJ]}{}^i&=0 \quad&\Longrightarrow\quad
 \xi^{\sss[k]}_{[I_1\cdots I_k]}{}^i&=0,\quad k\geq2,\\
 \theta^I&:=\tfrac1n\,\Gamma_i^{IJ}\xi'_J{}^i.
\end{alignat}

As before, neither super-commutativity nor nilpotence of any of the coordinates $(x^i|\xi'_I{}^i|\xi''_{[IJ]}{}^i|\cdots)$ is required by consistency of the supersymmetry algebra~(\ref{eSuSyH}) and the canonical commutation relations~(\ref{eCCRH}). This then permits the definition of this obstruction space, $Y$, to super-commutativity and nilpotence of $\wQ(X)$, and $\wQ(X)$ is naturally fibered over the obstruction space, $Y$. However, now already the straightforward generalization of Eq.~(\ref{ee0}),
\begin{equation}
 \eta^{ij}:=[x^i,x^j],
 \label{eeij}
\end{equation}
is a nontrivial obstruction to super-commutativity, and has played an important r\^ole in some recent studies in non-commutative field theory upon the requirement
\begin{equation}
 [x^i,x^j]~=~\Theta^{ij}=\textit{const}.,
 \label{eACSpT}
\end{equation}
which is easily accomplished using a straightforward generalization of~(\ref{eDisC}). The general setting describing obstructions to super-commutativity in all of $\wQ(X)$ is then evidently a straightforward generalization of anticommutivity~(\ref{eACSpT}) in spacetime.

Continuing in this way, define $Z$ to be the obstruction space of super-commutativity and nilpotence in the total space of the fibration of $\wQ(X)$ over $Y$, and this iterative telescoping procedure may be consistently continued indefinitely.

\section{The Comfortably Vast Superspace}
The result is an indefinitely and hierarchically telescoping tower of fibered spaces, $\mathscr{S}_N(X)$, starting with $\wQ(X)$, the local coordinates of which form a supermultiplet generated as the complete $Q$-orbit starting from spacetime coordinates, $x^i:X\to\mathbb{R}$. I should like to dub this the \raisebox{-1pt}{\includegraphics[height=2.3ex]{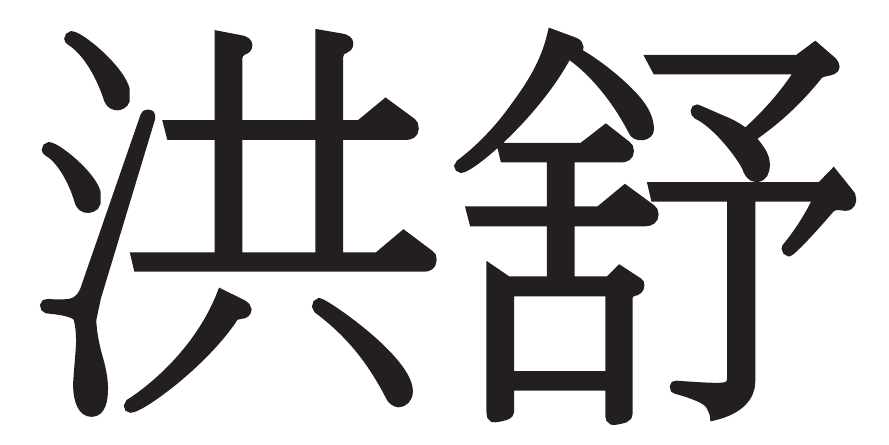}} (H{\'o}ng Sh{\=u}) superspace. The first of these characters translates as ``vast,'' while the second one can mean ``comfortable'', and $\mathscr{S}_N(X)$, as sketched out above, is certainly both (indefinitely) vast and so almost certainly comfortably sufficient to describe any supersymmetric theory ever needed. Serendipitously, this also happens to be the name bestowed upon me by Prof.~Shing Tung Yau, to whom I wish a happy 59/60$^\text{th}$ birthday!

\appendix
\section{Jacobi Identities and Permutation Symmetry}
 \label{a:JI}
The four Jacobi identities take the general form:
\begin{alignat}{3}
 &[B_1,B_2,B_3\}:&\qquad
 0&=\big[B_1,[B_2,B_3]\,\big]
   +\big[B_2,[B_3,B_1]\,\big]
   +\big[B_3,[B_1,B_2]\,\big]~, \label{eBBB}\\[2mm]
 &[B_1,B_2,F\}:&\quad
 0&=\big[B_1,[B_2,F]\,\big]
   +\big[B_2,[F,B_1]\,\big]
   +\big[F,[B_1,B_2]\,\big]~, \label{eBBF}\\[2mm]
 &[B,F_1,F_2\}:&\quad
 0&=\big[B,\{F_1,F_2\}\,\big]
   +\big\{F_1,[F_2,B]\,\big\}
   -\big\{F_2,[B,F_1]\,\big\}~, \label{eBFF}\\[2mm]
 &[F_1,F_2,F_3\}:&\quad
 0&=\big[F_1,\{F_2,F_3\}\,\big]
   +\big[F_2,\{F_3,F_1\}\,\big]
   +\big[F_3,\{F_1,F_2\}\,\big]~, \label{eFFF}
\end{alignat}
where the $B$'s and $F$'s represent (even) bosons and  (odd) fermions, respectively.
 These may be used to express any one of the summands in terms of the other two. For example, Eq.~(\ref{eBFF}) implies that:
\begin{equation}
 \big\{F_1,[F_2,B]\,\big\}
   =\big\{F_2,[B,F_1]\,\big\}
    -\big[B,\{F_1,F_2\}\,\big]
   =-\big\{F_2,[F_1,B]\,\big\}
    -\big[B,\{F_1,F_2\}\,\big]~.
  \label{eFFB}
\end{equation}
Indeed, the latter version may be used to define $F_1\big([F_2,B]\big)$, the superderivative application of $F_1$ upon the commutator $[F_2,B]$. In this way, the successive application of  the supersymmetry algebra elements reduce to iterative applications of the defining equations~(\ref{eSuSyH}).

Using~(\ref{eBBB})--(\ref{eFFF}) and their `derivatives' such as~(\ref{eFFB}), compute:
\begin{alignat}{3}
 [H,B]&=-[B,H]
  =-\tfrac1{2N}\delta^{IJ}\,\big[B,\{Q_I,Q_J\}\,\big]&
 &=\tfrac1{N}\delta^{IJ}\,\big\{Q_{(I},[Q_{J)},B]\,\big\}~.\label{eHB}\\
 [H,F]&=-[F,H]
  =-\tfrac1{2N}\delta^{IJ}\,\big[F,\{Q_I,Q_J\}\,\big]&
 &=\tfrac1{N}\delta^{IJ}\,\big[Q_{(I},\{Q_{J)},F\}\,\big]~.\label{eHF}
\end{alignat}
Then, for $F'_I:=[Q_I,B]$,
\begin{alignat}{3}
 \{Q_I,F'_J\}&
  =\{Q^\9_{(I},F'_{J)}\}+\{Q^\9_{[I},F'_{J]}\}&
 &=\delta_{IJ}\,[H,B]
   +\big\{Q_{[I},[Q_{J]},B]\,\big\}~. \label{eQQB}
\intertext{Similarly, for $B'_I:=\{Q_I,F\}$,}
 [Q_I,B'_J]&
  =[Q^\9_{(I},B'_{J)}]+[Q_{[I},B'_{J]}]&
 &=\delta_{IJ}\,[H,F]
   +\big[Q_{[I},\{Q_{J]},F\}\,\big]~. \label{eQQF}
\end{alignat}

We have also used the iterative construction of representations of the permutation group. So, for example:
\begin{align}
 A_{(I}B_{J)}&:=\tfrac12(A_IB_J+A_JB_I),\qquad\text{and}\qquad
 A_{[I}B_{J]}:=\tfrac12(A_IB_J-A_JB_I),\\
 A_{(I}B_JC_{K)}
  &:=\tfrac16(A_IB_JC_K+A_IB_KC_J+A_KB_IC_J+A_KB_JC_I+A_JB_KC_I+A_JB_IC_K),\\
 A_{[I}B_JC_{K]}
  &:=\tfrac16(A_IB_JC_K-A_IB_KC_J+A_KB_IC_J-A_KB_JC_I+A_JB_KC_I-A_JB_IC_K),\\[-1mm]
\noalign{\noindent so\vspace{-1mm}}
 A_IB_{[JK]}
  &~=\tfrac12A_I(B_JC_K-B_KC_J),\nonumber\\
  &~=A_{[I}B_JC_{K]}
    +\tfrac13\big((A_{(I}B_{J)}C_K-A_{(I}B_{K)}C_J)
                 +(A_{(I}B_JC_{K)}-A_{(I}B_KC_{J)})\big),\\[-1mm]
\noalign{\noindent where the first sumand is totally antisymmetric, and the second summand is the \includegraphics[height=2.3ex]{Pix/YT21.pdf} representation of the permutation group:}
 (ABC)_{(IJ)K}
  &:=\tfrac13\big((A_{(I}B_{J)}C_K-A_{(I}B_{K)}C_J)
                 +(A_{(I}B_JC_{K)}-A_{(I}B_KC_{J)})\big),\\[-1mm]
\noalign{\noindent and may be identified with the kernel of the antisymmetrization map:\vspace{1mm}}
  &~\ni\ker\big[V\otimes\wedge^2V \to \wedge^3V\big],
\end{align}
where $V$ is the linear vector space of $N$-tuples $(A_1,\cdots,A_N)$, $(B_1,\cdots,B_N)$ and $(C_1,\cdots,C_N)$.

\bigskip\paragraph{Acknowledgments:}\small
I should like to thank C.F.~Doran, M.G.~Faux, S.J.~Gates,~Jr., K.M.~Iga and G.D.~Landweber for correcting many of my misconceptions, and the Department of Energy for the generous support through the grant DE-FG02-94ER-40854.  Some Adinkras were drawn with the aid of the {\em Adinkramat\/}~\copyright\,2008 by G.~Landweber.

\clearpage
\small
\def\rasp{\leavevmode\raise.45ex\hbox{$\rhook$}}

\end{document}